\documentclass[twocolumn, twocolappendix]{aastex631}

\shorttitle{Measure Reionization with Fast Radio Bursts}
\shortauthors{Heimersheim et al.}
\usepackage{csquotes}
\usepackage{booktabs}
\usepackage{url}
\usepackage{amsmath}

\newcommand{\DM}[0]{\mathrm{DM}}
\newcommand{\obs}[0]{\mathrm{obs}}
\newcommand{\sh}[1]{{#1}} %\textcolor{red}

\begin{document}
%Custom cite alias
\defcitealias{Planck2018}{\textit{Planck}}
\defcitealias{CHIMEFRBCatalog1}{{CHIME/FRB catalog}}

\title{{W}hat it {T}akes to {M}easure {R}eionization with {F}ast {R}adio {B}ursts}

\author[0000-0001-9631-4212]{Stefan Heimersheim}
\correspondingauthor{Stefan Heimersheim}
\affiliation{Institute of Astronomy, University of Cambridge, Madingley Road, Cambridge CB3 0HA, UK}
\email{heimersheim@ast.cam.ac.uk}

\author[0000-0003-2138-5192]{Nina S. Sartorio}
\affiliation{Institute of Astronomy, University of Cambridge, Madingley Road, Cambridge CB3 0HA, UK}

\author[0000-0002-1369-633X]{Anastasia Fialkov}
\affiliation{Institute of Astronomy, University of Cambridge, Madingley Road, Cambridge CB3 0HA, UK}
\affiliation{Kavli Institute for Cosmology, Madingley Road, Cambridge CB3 0HA, UK}

\author[0000-0003-1301-966X]{Duncan R. Lorimer}
\affiliation{West Virginia University, Department of Physics and Astronomy, P. O. Box 6315, Morgantown, WV, USA}
\affiliation{Center for Gravitational Waves and Cosmology, West Virginia University, Chestnut Ridge Research Building, Morgantown, WV, USA}



\begin{abstract}
%250 word limit
Fast Radio Bursts (FRBs) are extra-galactic radio transients which exhibit a
distance-dependent dispersion of their signal, and thus can be used as cosmological probes.
In this article we, for the first time, apply a   model-independent approach to measure
reionization from synthetic FRB data assuming these signals are detected beyond redshift 5. This method allows us to constrain
the full shape of the reionization history as well as the CMB optical depth $\tau$
while avoiding the {problems} of commonly used model-based techniques.
100 localized FRBs, originating from redshifts {$5-15$},
could constrain (at 68\% confidence level) the CMB optical depth to within 11\%, and the midpoint of reionization to 4\%, surpassing current state-of-the-art CMB bounds and quasar limits. Owing to the higher numbers of expected
FRBs at lower redshifts, the $\tau$ constraints are asymmetric ($+14\%, -7\%$)
providing a much stronger lower limit.
Finally, we show
that the independent constraints on reionization from FRBs will improve limits  on other cosmological parameters such as
the amplitude of the power spectrum of primordial fluctuations.
\end{abstract}




\section{Introduction}
\label{sec:intro}
A type of radio transient signals, called Fast Radio Bursts (FRBs), is one of the unsolved mysteries in astronomy. The quest for understanding their origins is ongoing, and, since the discovery of the first FRB in 2007 \citep{Lorimer_2007}, a few hundreds such signals have been detected by the most sensitive radio telescopes.

These signals are significantly dispersed by the ionized medium distributed along the path between the sources and the observer. Since the observed dispersion, quantified by the dispersion measure (DM), is much larger than the contribution of our local environment, it is now known that the overwhelming majority of FRBs are of extra-galactic origin. For the bulk of the detected FRBs, a large fraction of the DM is due to the free electrons found in the intergalactic medium
\citep[IGM,][]{Xu_2015}, with a lesser contribution from the FRB host environment and the Milky Way.
\sh{Evolution} of DM \sh{with} the cosmological redshift of the FRB \sh{source} may shed light on properties of the IGM  \citep{Macquart_2020} %
and can be used to measure cosmological parameters
{\citep{Deng2014, Zhou2014, Gao2014, Yang_2016, Walters2018, Jaroszynski2019, Kumar_2019, Madhavacheril2019, Walters2019, Wu2020, Hagstotz2021}}.

Existing data \sh{suggest} that the number of FRBs traces the star formation rate  \citep[SFR,][]{Arcus2021,James:2021}, \sh{implying} that FRBs could exist beyond the current observational cutoff at  $z \sim 3$ \citep{2018Bhandari}.
\sh{The expected relation between the SFR and the detection rate of high-redshift  FRBs} is uncertain as both the FRB luminosity function and the nature of FRB progenitors are poorly understood and might change as  stellar populations evolve.
The recent detection of a Galactic FRB 200428 associated with magnetar SGR 1935+2154 \citep{Chime_2020, Bochenek_2020} suggests that at least some FRBs are related to the highly magnetized neutron stars. However, even if all FRBs were produced by magnetars, the ambiguity remains as there are multiple channels through which a magnetar could give rise to an FRB
\citep[see e.g.][]{Popov_2010arXiv, Petroff_2016,  Wadiasingh_2020}.

It is yet unknown when the first FRB-producing systems  might have formed. Theoretical models predict an early onset of star formation with the first stars in the observable Universe appearing just a few tens of million years after the Big Bang \citep{Naoz:2006}. \sh{Additionally}, observations of high-redshift galaxies find evolved and luminous objects  including a  bright galaxy at $z\sim11$ detected by the Hubble Space Telescope  \citep[HST,][]{Oesch:2016} and a metal-rich galaxy at   $z\sim9.1$ observed by the Atacama Large Millimeter/submillimeter Array, HST and {\it Spitzer} whose existence requires star formation to begin at  $z\gtrsim15$ \citep{Hashimoto:2018}.
Thus, if the  correlation between SFR and the number of FRBs holds for the entirety of cosmic history,
\sh{we can expect a multitude of FRBs produced in  high-redshift galaxies \citep[][]{Fialkov:2017, Hashimoto2008.00007}.}  However,  the number of these high-redshift events is expected to depend on the highly uncertain properties of the first stars and galaxies   \citep[e.g. the initial mass function of the first generation of stars,][]{Klessen:2019}.

From the observational point of view, detecting high redshift FRBs
requires both, sufficient telescope sensitivity and a capability to detect highly dispersed events.
Some existing telescopes already have these specifications, including the Five-hundred-meter Aperture {Spherical radio} Telescope {(FAST)} which could detect the brightest FRBs up to $z \sim 10$ \citep{Zhang:2018} and the Green Bank Telescope carrying out a high-sensitivity search using the GREENBURST experiment \citep{Agarwal:2020}.
Next generation telescopes like the Square Kilometre Array (SKA) will be able to detect multiple FRBs per day from high redshifts, potentially originating in the first billion years of cosmic history \citep{Fialkov_2016, Hashimoto2020SFR}.

If detected, high redshift FRBs will be extremely rewarding in terms of the science they deliver. First, either detection or non-detection of these events would probe progenitor theories \sh{and act} as an indirect {constraint on} star formation at high redshifts. Second, high-redshift FRBs are expected to be sensitive cosmological probes
especially if accurate measurement of host-galaxy redshift via source localization are available.
\sh{Such capability already exists for the low-redshift FRBs observed with the Australian Square Kilometre Array Pathfinder \citep[ASKAP,][]{Bannister2019}.}  Theoretical predictions show that a population of FRBs at $z\sim3-4$ could be used to probe the second Helium  reionization happening around redshift $z\approx 3.5$
{\citep[][]{Zheng2014, Linder2020, Bhattacharya2020}}, while signals from $z>5$ would primarily trace the ionization state of Hydrogen atoms, and, consequently, could be used to probe the Epoch of Reionization \citep[EoR;][]{Fialkov_2016,Dai2020,Beniamini2020,Zhang2021,HashimotoReioH2021,Pagano2021}.
Such measurements would help constraining \sh{the history of reionization} and the optical depth of the Cosmic Microwave Background (CMB), $\tau$, to a  higher accuracy than what is currently
achievable \citep{Fialkov_2016, Dai2020, Beniamini2020, Zhang2021}. 

Current best limits on  $\tau$ have a large uncertainty
 \citep[$\sim\,13\%$][]{Planck2018}
and represent a hurdle for precision cosmology. These limits are derived from the CMB which is affected by the EoR as photons scatter off the ionized gas. The signature of the EoR in the CMB is  degenerate with the initial amplitude of
curvature perturbations $A_s$
\citep{Hazra2018}, and, thus, the poor precision in $\tau$  also degrades the estimates of the primordial power spectrum.
Another quantity that suffers from the poor constraint of $\tau$ is the sum of neutrino masses, $\Sigma m_\nu$,
extracted from a joint analysis of the CMB and Large Scale Structure \citep{Brinckmann2019}.
Independent knowledge of the ionization history will also be important for 21-cm cosmology, breaking degeneracies between the ionization and thermal histories and allowing a measurement of the IGM temperature from the 21-cm signal  \citep{Fialkov:2017}.
A more accurate measurement of the EoR history is \sh{therefore crucial} to improve our fundamental understanding of the Universe.

Existing works constraining Hydrogen reionization based on localized FRBs \citep{Fialkov_2016, Beniamini2020, Zhang2021, HashimotoReioH2021, Pagano2021}
make {different} assumptions that {can} either
lead to biases in the 
results or limit the constraining power of the data, {especially
as our understanding and observations of FRBs and the IGM get better.}
Early studies \citep{Fialkov_2016} point out the relation between $\mathrm{DM}(z)$ and $\tau(z)$,
but do not consider multiple measurements distributed over redshift {or} the impact of the redshift-dependence of the ionized fraction ($x_i(z)$, also referred to here as the \enquote{shape} of the EoR history).
Some recent analyses \citep[one {of the methods adopted in}][]{Beniamini2020,HashimotoReioH2021}
consider constraining  the ionized fraction by
grouping FRBs into redshift
bins and {using the average $\mathrm{DM}$ in each bin.}
{This allows measuring $x_i$ between the bins but {loses} %
information about the  redshifts and DMs  of individual FRBs, thus,} limiting the resolution with
which the reionization history can be traced.
Furthermore, in the absence of a functional form for \sh{$x_i(z)$}, this approach does not permit imposing global relationships such as monotonic \sh{evolution with} redshift,
\sh{or {integrating over $x_i(z)$} to compute} $\tau$.
A different approach is to postulate a specific shape of the reionization history, either using the popular \textit{tanh}
function \citep{Zhang2021} or a simulation-based form \citep{Pagano2021}. Because the actual EoR history is unknown, \sh{however},
assuming a specific history
{can lead to a significant offset in the results as we discuss}
in Section \ref{sec:results}.

In this paper we, for the first time, use a free-form approach  to estimate how well the EoR history can be probed with high-redshift  FRBs (from $z>5$) assuming these events are observed and localized with the future sensitive telescopes (e.g. the SKA).
We use a free-form reionization parameterization \enquote{FlexKnot} \citep{Millea2018}, which allows us to constrain the full shape of the EoR history thus yielding robust constraints on the optical depth $\tau$. As this method does not assume a specific functional form for the reionization history,
{we can obtain a model-independent inference of $\tau$ and $x_i(z)$.}%

This paper is organized as follows: In the next Section \ref{sec:DMmethods} we  briefly review different contributions to the observed FRB dispersion measure and discuss the  associated uncertainties. In  Section \ref{sec:methodsData} we describe the synthetic data set of high redshift FRBs, produced for this work. In Section \ref{sec:FlexKnot}  we present the model-independent method which is used to parameterize the EoR history. In Section \ref{sec:inf} we outline the likelihood and inference procedure. The results are discussed in Section \ref{sec:results} where we show the FRB constraints on the full reionization history as well as on the values of the optical depth $\tau$ and other cosmological parameters. Finally we summarize our findings in Section \ref{sec:conclusions}.

\section{Dispersion Measure}
\label{sec:DMmethods}

The DM of FRBs is commonly decomposed into three parts:
contribution from free electrons in the IGM,
the FRB host galaxy and the Milky Way
\begin{gather}
    \DM = \DM^{\rm IGM} + \DM^{\rm host} + \DM^{\rm MW}.
    \label{eq:DM_contributions}
\end{gather}
We compute each one of these  contributions as specified below, model the uncertainty of each component and, in addition, account for an observational errors. Approximating the components as statistically independent Gaussian random variables,
we add the uncertainties in quadrature
\begin{gather}
    \sigma_\DM = \sqrt{\left(\sigma^{\rm IGM}_{\rm DM}\right)^2 + \left(\sigma_\DM^{\rm host}\right)^2
    + \left(\sigma_\DM^{\rm MW}\right)^2
     + \left(\sigma_\DM^\mathrm{obs}\right)^2}.
     \label{Eq:sigmaDM}
\end{gather}

\subsection{IGM contribution}
\label{subsec:DMIGM}
The largest and the most important contribution to the dispersion of high redshift FRBs
is $\DM^{\rm IGM}$. Owing to the fluctuations in electron density along the line of sight,
this quantity is stochastic with 
the distribution approaching a Gaussian \citep{Macquart_2020},
as demonstrated by studies of the DM derived from numerical simulations \citep{Jaroszynski2019, Batten2020, Zhang2021}.
For FRBs originating from redshift $z$, the mean IGM contribution is given by
considering a homogeneous distribution of matter (electrons)
\begin{gather}
    \overline{\DM}^{\rm IGM}(z) = \int_0^{z} \frac{c \bar n_e(z')}{H(z') (1+z')^2} \mathrm{d} z'
    \label{eq:DM}
\end{gather}
where $\bar n_e(z')$ is the mean number density of free
electrons
at redshift $z'$, $c$ the speed of light and $H(z')$
is the Hubble expansion rate.
In order to explicitly compare the contributions due to reionization and other cosmological parameters we can
separate the terms as
\begin{align}
\overline{\DM}^{\rm IGM}(z) = \,& \frac{3c}{8\pi G m_p}\int_0^{z}
    \underbrace{{(1+f_{\rm He}) x_i(z) + f_{\rm He} x_i^{\rm HeII}(z)} }_{\mathrm{Reionization\ history}}  \nonumber \\
    &\times\underbrace{\frac{\Omega_b H_0 (1+z')}{\sqrt{\Omega_m(1+z')^3+(1-\Omega_m)}}}_{\mathrm{Background\ cosmology}} \label{eq:DMhomogcosmo}
    \,\mathrm{d} z'
\end{align}
where we wrote $H(z)$ and $\bar n_e(z)$ in terms of the Hubble constant $H_0$,
the matter density parameter $\Omega_m$, the \sh{physical baryon density parameter} $\Omega_b H_0^2$,
the gravitational constant $G$, the proton mass $m_p$,
and the number of Helium relative to Hydrogen atoms $f_{\rm He}$.
We also introduced the ionized fractions $x_i^{\rm HeII}$ for the \sh{second} Helium reionization (occuring at $z = 3-4$), and $x_i(z)$ quantifying the Hydrogen (and \sh{first} Helium) reionization.
At the high redshifts of interest ($z>5$), the integrand only depends on $x_i(z)$
and a factor proportional to $\Omega_b H_0/\Omega_m^{-0.5}$.

{Note that FRBs actually probe the electron density in the IGM, which is slightly lower than the mean cosmic density $\bar n_e(z)$ due to the contribution of  electrons in stars and black holes. This discrepancy is usually mitigated by introducing a factor $f_{\rm IGM}$, the fraction of electrons in the IGM,  which plays an important role in cosmological studies  with  low-redshift FRBs (and causes a fixed offset of the DM value for $z>5$ FRBs). However, at high redshifts this fraction
is close to unity, as was shown by  \citet{Takahashi2020} in a simulation-based study. The authors find $f_{\rm IGM}>0.98$ at $z>5$ and $f_{\rm IGM}>0.996$ at $z>10$. Assuming  $f_{\rm IGM} = 1$ at $z>5$ leads to a negligible error ($< 1\%$) on $\tau$.}

We estimate the uncertainty on $\DM^{\rm IGM}$ using the    {popular low-redshift ($z< 3$)}  relation {\citep{Kumar_2019}}
\begin{gather}
    \sigma^{\rm IGM}_{\rm DM}(z) = \left(0.2/\sqrt{z}\right) \overline{\DM}^{\rm IGM}
    \label{eq:sigIGM}
\end{gather}
derived from simulations {\citep{Jaroszynski2020} which we extend  to higher redshifts. 
We extrapolate the relation to the maximum uncertainty
at $z_{\rm peak} \approx 6.5$ 
and then assume this value for $z>z_{\rm peak}$, i.e. we use Eq.
\ref{eq:sigIGM} with
$\sigma^{\rm IGM}_{\rm DM}(z>z_{\rm peak}) = \sigma^{\rm IGM}_{\rm DM}(z_{\rm peak})$.
Such a treatment has been
employed previously by \citet{HashimotoReioH2021}, albeit with a fixed $z_{\rm peak} = 6$. We note that if Eq.   \ref{eq:sigIGM} were revised in the future with new observations or simulations becoming available, the numerical value of the uncertainty in DM would, naturally, be affected. However, such a change would not have an impact on a comparative analysis of statistical methods, such as the one laid out here.}

\subsection{Galactic contributions}
The contributions to the total DM by the FRB host galaxy and the Milky Way are expected to be small compared to that of the IGM, which is $\gtrsim 4500\,\mathrm{pc\,cm^{-3}}$ for
the bursts produced at high redshifts $z>5$.

The host contribution is not well determined
\citep[e.g.][ found $55\lesssim \DM_{\rm host}\lesssim225$]{Tendulkar2017}, and different values are 
used in the literature.
Conservatively, we adopt a rather large uncertainty of $\sigma_\DM^{\rm host} = 100\,\mathrm{pc\,cm^{-3}}$ \citep[following][]{Dai2020} in the host galaxy rest frame.

The Milky Way contribution \citep[$\lesssim 500\,\mathrm{pc\,cm^{-3}}$ for
samples in the \citetalias{CHIMEFRBCatalog1},][]{CHIMEFRBCatalog1}
can be estimated from Galactic electron models for every line of sight 
\citep{Cordes_2002arXiv, Yao2017} and subtracted from the DM measurement, thus it will not have a large impact when estimating the cosmological contribution of the high-redshift FRBs. 
We assume a $20\,\mathrm{pc\,cm^{-3}}$ modelling uncertainty, motivated by the
average difference between {NE2001} \citet{Cordes_2002arXiv} and {YMW217} \citet{Yao2017} models in the \citetalias{CHIMEFRBCatalog1}.
To account for the uncertain Milky Way halo contribution
\citep{Yamasaki2020, Keating2020, Das2021} 
we add an \sh{error of} $50\,\mathrm{pc\,cm^{-3}}$. Thus, with the errors combined in quadrature we obtain
$\sigma_\DM^{\rm MW} \approx 54\,\mathrm{pc\,cm^{-3}}$.

\subsection{Measurement uncertainty}
The \sh{final uncertainty on the DM} is the measurement error $\sigma_\DM^{\rm obs}$, which quantifies the accuracy in the FRB arrival time at different frequencies. To calculate the uncertainty we follow \citet{Lorimer2013MNRAS.436L...5L}  which gives
\begin{align}
	\sigma_\DM^{\rm obs} &= \frac{\nu^3 T_\mathrm{sys}}{kGs} \sqrt{\frac{W}{2\Delta \nu^3}}
	\end{align}
where $\nu$ is the observed frequency, $\Delta\nu$ is the  bandwidth, $W$ is the observed pulse width, $G$  is the instrumental gain,  $T_{\rm sys}$ is the system temperature, $s$ is the peak spectral flux density of the signal, and $k$ is a constant $k=4.15 \mathrm{\,GHz^2\, ms\, cm^3\,pc^{-1}}$ \citep[e.g.][]{Kulkarni2020arXiv}.
In other words, $\sigma_\DM^{\rm obs}$ is  mainly determined by  the observed pulse width  and the signal to noise ratio depending on the  peak spectral flux density of the signal
as well as on the instrumental
parameters \citep[][]{2016era..book.....C}.

The observed pulse width is set by two factors: the intrinsic restframe width of an FRB, $W_{\rm int}$, taken to be  $1\,\mathrm{ms}$, and a contribution  due to dispersion smearing (the latter is small, $\leq 10\%$) and is given by 
\begin{align}
	W &= \sqrt{[W_{\rm int} (1+z)]^2 + (\DM\,k  \Delta \nu/N_{\rm chan}\nu^3)^2}
	\end{align}
where $z$ is the redshift of the FRB host galaxy, $N_{\rm chan}$ is the number of channels and $\DM$ is the total dispersion measure.\footnote{The total
dispersion measure is dominated by the IGM contribution ($4500-6000\,\mathrm{pc\,cm^{-3}}$,
depending on redshift and the reionization history) and its scatter. To make sure the other contributions have a 
negligible effect on the dispersion smearing we add $(500+50+200)\,\mathrm{pc\,cm^{-3}}$
as conservative estimates of Milky Way ($\lesssim 500\,\mathrm{pc\,cm^{-3}}$ for
the sample in the \citetalias{CHIMEFRBCatalog1}), halo \citep{Yamasaki2020}, and
host galaxy \citep{Tendulkar2017} contributions, respectively.}
Note that the  sampling and scattering times are $\lesssim 10\,\mathrm{ms}$ and, thus, are negligible for high redshift FRBs.

For our analysis we use the most distant  (at the time of writing) localized FRB, FRB180924 \citep[][]{Bannister2019}, as a proxy for the  peak spectral flux density $s$. We scale the peak flux value to higher host galaxy redshifts accounting for the luminosity distance and the spectral shape and assuming spectral index $\alpha=-1.5$ compatible with the recent  results of CHIME/FRB \citep{CHIMEFRBCatalog1}. For comparison, FRB180924 has a higher intrinsic luminosity than about 75\% of the \citetalias{CHIMEFRBCatalog1} samples.
The instrument parameters are taken to be similar to FAST \citep{Nan2011} with $G=15\mathrm{\,K\,Jy^{-1}}$,  $T_{\rm sys}=35\mathrm{\,K}$,  $\nu=1.4\mathrm{\,GHz}$,   $\Delta\nu = 400\mathrm{\,MHz}$, and $N_{\rm chan}{=}\,4096$.
{For completeness we take this uncertainty into account but its effect is mostly negligible, contributing $\sigma^{\rm obs}_{\rm DM}\lesssim 100\,\mathrm{pc\,cm^{-3}}$.}

\section{Synthetic data set}
\label{sec:methodsData}

We create a synthetic data set of high redshift FRBs in order to demonstrate the potential of these sources in constraining the EoR. We generate two data sets, one  with 100 and the other with 1,000 localized FRBs.
Each data set consists of the host galaxy redshifts $z^\mathrm{obs}$ generated from a distribution proportional to SFR, and dispersion measures $\mathrm{DM}^\mathrm{obs}$ calculated following the prescription outlined in Section \ref{sec:DMmethods}.

\begin{figure}
    \centering
    \includegraphics[width=.5\textwidth]{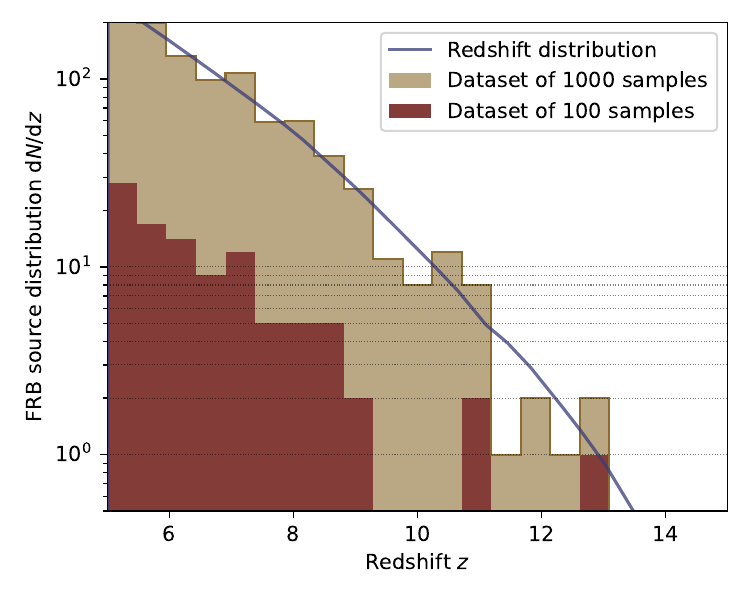}
    \caption{{Source redshift samples for 100 and 1,000 FRBs used in the analysis (histograms), compared to the analytical distribution (line plot) following SFR. The source redshifts are sampled from
    this distribution using a simple Monte Carlo rejection sampling.}}
    \label{fig:frbdistribution}
\end{figure}

As we are interested in the EoR constraints, we consider host galaxies in the redshift range
$z^{\rm true}= 5-15$. We draw these true source redshifts from a distribution scaling with
the cosmic SFR, adopting a fit from \citet{Behroozi2019}, {as shown in Figure \ref{fig:frbdistribution}}. This choice is rather pessimistic, {allowing for few $z>10$ FRBs} as the adopted SFR drops steeply at high redshifts.
{Concretely we find that we sample}
85\% of FRB sources between redshift
$z^{\rm true}=5$ and 8, 12\% between 8 and 10, 3\% between 10 and 15, and
no sources beyond $z^{\rm true}>15$.
However, as mentioned in Section \ref{sec:intro}, the high-redshift star formation  is not well constrained and a much higher SFR during the early stages of the EoR is plausible.   A more efficient star formation at high redshifts  would result in a larger number of FRBs and, thus, even stronger EoR constraints than what we report here, {while a lower SFR would imply less FRBs and, consequently, weaker constraints}. The method presented in this paper is generic and can be applied to an arbitrary SFR.

After generating the true host galaxy redshifts we simulate the observed redshifts $z^\obs$,
assuming an accuracy of 10\% (such as provided by spectra of host galaxies from the James Webb Space Telescope), and draw  $z^\obs$ from a Gaussian
distribution $\mathcal{N}(\mu_z=z^\mathrm{true}, \sigma_z=0.1 z^\mathrm{true})$.

Finally we simulate the corresponding synthetic DM observations, $\mathrm{DM}^\mathrm{obs}$,
by adding the IGM, Milky Way, and host galaxy contributions,
and taking the associated
uncertainties as well as the measurement errors  into account, in agreement with Section \ref{sec:DMmethods}.
The IGM contribution is computed based on the true source redshift $z^{\rm true}$,
and using the reionization history from \citet{Kulkarni2019}, compatible with current
observational constraints on the EoR, and the cosmological parameter values
$\Omega_{m}=0.3144$, $\Omega_b=0.04938$,
and  $H_0=67.32\,\mathrm{km\,s^{-1}\,Mpc^{-1}}$, 
compatible with \textit{Planck} \citep{Planck2018}.
We point out that these inputs are only used to generate the synthetic data sets. Our inference procedure, which is described in Sections \ref{sec:FlexKnot} and \ref{sec:inf}, is completely blind to both the cosmological  parameters and the input EoR history.

{To check that our conclusions do not depend on the specific realization of the synthetic data, we re-run the analysis with two more random realizations of redshift and DM samples. Naturally, the exact numerical values  vary from one realization  to another (with the strongest impact coming from the number of FRBs observed at $z>10$), but all runs give comparable constraints
and posterior bounds consistent
with the true (input) value. In 18 out of 24
computed quantities (8 per run, as in Table \ref{tab:resultstable}) the true (input)
value lies within the 68\% confidence interval,
and in all cases lies within the 95\% interval.}

\section{Parameterization of the EoR history}
\label{sec:FlexKnot}

For either a synthetic or a real sample of high-redshift FRBs, the relation between the measured $z^\mathrm{obs}$ and $\mathrm{DM}^\mathrm{obs}$   depends on the ({a priory} unknown) values of the cosmological parameters  $\Omega_b, \Omega_m, H_0$ and, crucially, on the full shape of the reionization history, $x_i(z)$, via the IGM contribution to DM. Therefore, these quantities can be extracted from a sample of high-redshift FRBs. In order to build an inference pipeline, which we describe in Section \ref{sec:inf}, we, first, need to create a
parameterization of the %
theoretical dispersion measure, $\DM^{\rm model}(\theta,z')$ following  Equations \ref{eq:DM_contributions} to \ref{eq:DMhomogcosmo}, as a function of the true redshift of the source, $z'$, and a vector of model parameters $\theta$, which includes the cosmological parameters and the full information about the EoR.  

The theoretical dependence of the dispersion measure on the cosmological parameters can straightforwardly be included in the analysis. We take Gaussian priors on the parameters $\Omega_b H_0=3.31\pm0.017$ and $\Omega_m=0.311\pm0.0056$, based on \citetalias{Planck2018}.
However, expressing the dependence of  $\DM^{\rm model}(\theta,z')$ on $x_i(z)$ is a more challenging task, as we have to avoid introducing a specific functional form or model.\footnote{Choosing specific functional forms would lead to biases in the results, as
discussed in the Section \ref{sec:intro} and illustrated in Section \ref{subsec:opticaldepth}.}
To this end  we use {a modified version of} the free-form FlexKnot method \citep{Millea2018, Vazquez2012}.

\begin{figure}
    \centering
    \includegraphics[width=.5\textwidth, trim={0.5cm 1.51cm 0 0},clip]{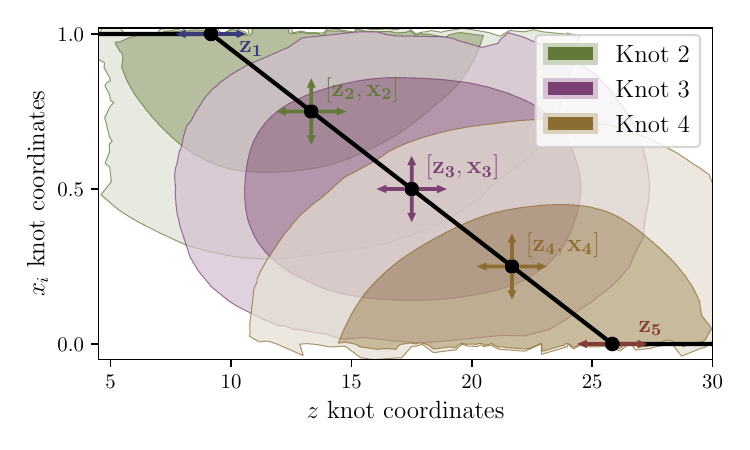}
    \includegraphics[width=.5\textwidth, trim={0.5cm .5cm 0 0.2cm},clip]{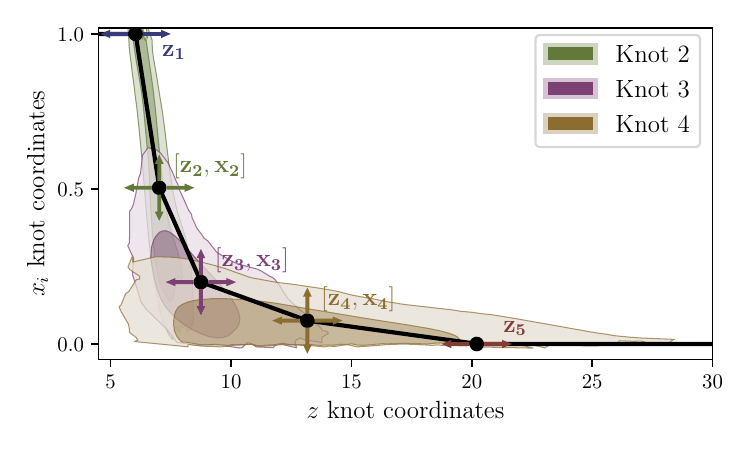}
    \caption{{FlexKnot reionization history with $N=5$ knots, parameterized by the coordinates of the interpolation knots [$z_1,x_{i,1}{=}1$], [$z_2, x_{i,2}$], [$z_3, x_{i,3}$], [$z_4, x_{i,4}$], [$z_5, x_{i,5}{=}0$].
    The points connected by the
    solid line show the knots and interpolated FlexKnot reionization
    history, the arrows indicate the possible movement of the knots.
    The contours show the positions (for $N=5$, 1,000 FRBs) of the
    intermediate knots, showing how the knots move from priors (upper panel) to the
    posteriors (lower panel).
    Note that the knots tend to cluster at lower redshifts, in the posterior. This is due to larger variation of $x_i(z)$ as well as more FRBs being observed at lower redshifts;
    The knots can more precisely constrain the shape of reionization if they are located at low $z$.
    }
    }
    \label{fig:flexknotillustration}
\end{figure}

The  FlexKnot method allows us to scan the space of all possible theoretical EoR histories by parameterizing the  ionized fraction with  $N$ points
(referred to as \enquote{knots}); Figure \ref{fig:flexknotillustration} shows an example for $N=5$.
These knots can move in order to  explore a wide range of plausible reionization histories and determine which histories are compatible with data (either synthetic, as in our case, or future real observations).
Each knot is parameterized by a parameter pair $[z_n,~x_{i,n}]$ corresponding to its redshift  and the value of the ionized fraction 
 at this redshift. Both of these parameters are varied, thus allowing the knot to move freely, {in both, $z$ and $x_i$.
 This typically leads to most knots moving towards
 lower redshifts where $x_i(z)$ varies more strongly to achieve
 a better fit.}
The  reionization history $x_i(z)$ at any arbitrary redshift is then given by the linear interpolation between these knots \citep{Handley2019Infl}, as illustrated   in Figure \ref{fig:flexknotillustration} (solid line).
In principle, knots can assume any value, but here we require the reionization history to be monotonic, such that the IGM can only become more ionized with time. \sh{We also require
the first knot (at the lowest redshift) to assume the value of  $x_{i,1}=1$, thus requiring reionization to be completed by $z_1$, and the last knot
to have $x_{i,N}=0$, implying that the universe is still fully neutral at $z_N$. %
We allow the redshift locations of the knots to assume any positions consistent with observations, that is, reionization has to be completed by $z=5$ \citep{McGreer2015} and begin no earlier than $z=30$ \citep[as the first stellar populations are expected to form then, e.g. ][]{Klessen:2019}.}

In principle, other model independent techniques, such as  principle components analysis \citep[PCA,][]{Hu2003,Heinrich2018} or interpolation between values at fixed redshifts   \citep[\textit{Poly-reion},][]{Hazra2017}, could be used to parameterize the EoR history instead of the FlexKnot method. 
What makes  FlexKnot stand out compared to the other techniques is not only that both the redshifts and the ionized fractions of the knots are free to vary, but also that the number of knots $N$ is a free parameter.
{This is equivalent to combining models with different numbers of knots $N$ and weighing them by their Bayesian evidence. High values of $N$ allow the fit to adapt to arbitrarily complex
EoR histories while, at the same time,}
histories with more knots (higher $N$) than warranted by the data will be down-weighted by a lower Bayesian evidence.

By design, weighing with respect to Bayesian evidence introduces a slight preference towards simpler reionization histories described with fewer interpolation knots
(and, thus, having less variation in the slope). For example, if the data mostly consist of low-redshift FRBs with almost no high redshift sources, then the strongest constraints on the ionization history are at low redshifts.
Such data could be compatible with either a simple late  reionization history that happens entirely within the redshift range probed by the FRBs, or with a history that has a
small high-redshift tail in the redshift range that is not constrained by the data. Because the former model is simpler (i.e. it requires less knots and parameters
to describe), it is preferred by the algorithm. As we will see in Section \ref{subsec:reionhist} this preference for simpler models is not a problem in practice. Even when the maximum posterior indicates a late reionization, the confidence intervals allow for a wider range of reionization histories, including the ones  that are more complicated.

\newpage
\section{Parameter inference}
\label{sec:inf}

In this section  we  describe the inference pipeline that is used to constrain reionization history
as well as those cosmological parameters  that affect the dispersion measure of FRBs.

For a given set of model parameters $\theta$, including  the FlexKnot parameters [$z_n, x_{i,n}$] 
for knots $n=1...N$ and the cosmological parameters $\Omega_b H_0$ and $\Omega_m$, we
can calculate the theoretical dispersion measure relation, $\DM^{\rm model}(\theta,z')$, as a function of $z'$.
\sh{We compare this model to a  set of observations, $d_i=\{\DM_i^\obs, z_i^\obs\}$, taken from the synthetic catalogues (Section \ref{sec:methodsData}).  The likelihood} 
of observing the $i$th FRB   from a source at true redshift $z'$ 
with the expected dispersion measure  $\DM^{\rm model}(\theta,z')$ is
\begin{align}
    P\left(d_i|\theta, z'\right) = \,&\mathcal{N}\left(\frac{z_i^\obs-z'}{\sigma_z(z')}\right)\\
           &\times\mathcal{N}\left(\frac{\DM_i^\obs-\DM^{\rm model}(\theta,z')}{\sigma_\DM\left(z', \DM^{\rm model}(\theta,z')\right)}\right),
\end{align}
\sh{where we assumed Normal distributions for the uncertainties in both the redshift  and $\DM$.
Next, we marginalize over the true source redshift $z'$, averaging
over possible values weighted by the likelihood and its prior probability distribution, $P(z')$, which we assume to be $P(z') = z e^{-z}$ \citep{Zhou2014}.\footnote{In our calculation we adopted $P(z')$ from  \citet{Zhou2014}. However, in reality this probability is unknown. We can quantify the error introduced by {using an approximate prior}
as we know the actual redshift distribution of the synthetic data (it follows the SFR).
We find that the difference between using the true redshift prior (proportional to SFR)
and the approximation \citep[][]{Zhou2014} is small compared to other contributions,
e.g. the resulting error in the value of the CMB optical depth is only $\Delta\tau \lesssim 2$\%.}}
Finally, we multiply the independent FRB observations to obtain the likelihood for the
entire data set $D=\{d_i\}$:
\begin{gather}
    P(D|\theta) = \prod_{i} \int P(d_i|\theta, z') P(z') \,\mathrm{d} z'.
\end{gather}

We use {the Nested Sampling \citep{Skilling2004}} algorithm to derive the posterior distributions of both, the cosmological parameters and the reionization history. \sh{We sample {these} parameters assuming} flat priors on the positions of the $N$ knots  [$z_n, x_{i,n}$]  (between {5 and 30, and 0 and 1}, respectively),
and Gaussian priors on the cosmological parameters $\Omega_b H_0=3.31\pm0.017$ and $\Omega_m=0.311\pm0.0056$, based on \citetalias{Planck2018}. {Note that these Planck priors do not influence the $\tau$-posterior.}

The uniform FlexKnot priors imply 
non-uniform priors on derived parameters such as the midpoint of reionization
$z_{\rm mid}$ or the optical depth $\tau$ \citep{Millea2018}. When deriving constraints
on these parameters we correct for this effect using maximum entropy priors \citep[following][]{Handley2019}.
\label{paragraph:priorcorrection}

{The number of knots, $N$, is a free parameter in our method.
We use a simple  implementation, running the Nested Sampling algorithm individually for each value
of $N$ and then combining the runs. This method is equivalent to varying $N$ and all parameters simultaneously. In total we}
run ten FlexKnot models, from $N{=}\,2$ knots (with only beginning and end of reionization defined) to $N_\mathrm{max}{=}\,11$. We do not consider more complex models for this data set as for {$N\geq11$ the evidence is small} (less than 1\% of maximum evidence) and the effect on the results is negligible.
We find that, in the case of just 100 FRBs, the simplest model (2 knots) has the highest evidence $Z_1=1\pm0.24$  with 3 and 4 knot models yielding lower  evidences, $Z_3=0.57\pm0.13$ and $Z_4=0.35\pm0.08$, respectively. For 1,000 FRBs, models with 4 knots achieve the highest evidence.%

The FlexKnot method is very versatile and could easily allow us to incorporate additional constraints on the EoR history from the CMB data by taking into account the effect of $x_i(z)$ on the CMB power spectrum, as well as  from other tracers \citep[such as quasars,][however, adding such constraints is beyond the scope of this paper]{McGreer2015}.
Here, as a proof of concept, we make use of this flexibility by adding 
the constraint on the contribution of the high redshift universe between $z=15$ and 30 to the total optical depth, $\tau_{15,30}<0.019$ \citep[95\% confidence,][]{Heinrich2021}.
Note the differentiation between the total $\tau$ constraint,
and the partial $\tau_{15,30}$ limit: as the CMB experiments are more sensitive to ionization at
high redshift, constraints on  $\tau_{15,30}$ can
be derived in addition to the total $\tau$ constraint
\citep[see e.g.][for similar recent constraints]{Heinrich2018, Millea2018, Planck2018}.

\section{Results}
\label{sec:results}

Finally we apply our method to the synthetic FRB data sets. We make use of the flexibility of the FlexKnot method, which provides the complete reionization history for every sample of parameters, to derive constraints on some of the characteristic stages of reionization as well as on cosmological parameters $\tau$ and $A_s$.

\subsection{Reionzation history}
\label{subsec:reionhist}
We first focus on reionization history and demonstrate that strong constraints on the evolution of the ionized fraction with redshift, $x_i(z)$, can be achieved. To this end, we marginalize over individual knot positions and the relevant cosmological parameters (density parameter of matter $\Omega_m$ and baryons $\Omega_b$, and the Hubble constant $H_0$). We find that 1,000 FRBs provide
good constraints on the shape of reionization history parameterized with 4 interpolation
knots (maximum evidence at $N{=}\,4$), while the smaller 100 FRB data set is less sensitive
to the details   (maximum evidence for the simple $N{=}\,2$ model).

\begin{figure*}[ht]
    \centering
    \includegraphics[width=\textwidth]{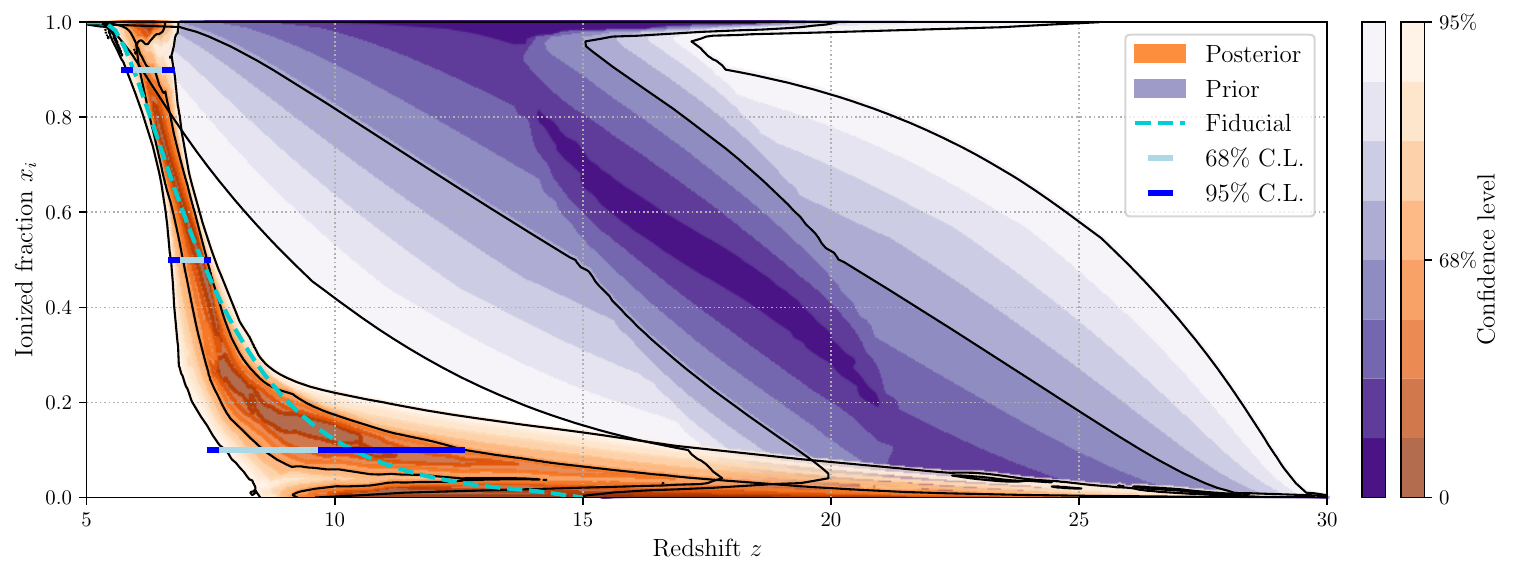}
    \caption{FRB constraints on the full reionization history.  We show the posterior (orange) and prior (purple) contours of the reionization history $x_i(z)$
    derived from 1,000 localized FRBs and including the additional constraint $\tau_{15,30}<0.019$   \citep{Heinrich2021}.
    The constraints are shown as iso-probability contours for $x_i(z)$ at any given redshift, with the confidence level
    indicated by the color (see colorbars), with 68\% and 95\% confidence contours marked as solid lines.
    Additionally, the error bars  show the uncertainties in the redshifts of start ($x_i=0.1$), midpoint ($x_i=0.5$) and end of reionization ($x_i=0.9$)  as 68\% (light blue) and 95\% (dark blue) confidence intervals.
    The dashed cyan line indicates the input (fiducial) reionization history \citep[adopted from][]{Kulkarni2019} that was used to generate the synthetic FRB data.
    }
    \label{fig:flexknotxi}
\end{figure*}

The constraints on the ionized fraction $x_i(z)$
obtained from 1,000 FRB observations, and additionally using the constraint on $\tau_{15,30}$,
 are demonstrated in Figure \ref{fig:flexknotxi}. The posterior and prior distributions of $x_i(z)$ are shown by the orange and purple contours respectively, with 68\% and 95\% confidence intervals marked by the solid lines.
The prior distribution  follows from assuming a monotonic reionization history and imposing a uniform prior on each individual knot position. This leads to a weak, non-uniform, prior on the interpolated value  of $x_i(z)$  at any given redshift, as indicated by the contours. 
The posterior provides tight constraints on the reionization history,
and encompasses the input reionization history that was used to generate the data (dashed cyan line in the figure) within the 95\% confidence intervals.%

The synthetic FRB data can constrain the  reionization history especially well at $z\lesssim 10$ where the vast majority (97\%) of the FRB
sources in our catalogues are located. The fact  that there is only a small number of FRBs at higher redshifts (owing to the steep drop in the adopted SFR, {with no FRBs beyond $z>14$}) results in  a weaker
constraint at $z{=}\,10-30$. The high-$z$ constraints
are, to a large part, driven by (i) the enforced monotonicity of the function, (ii) the preference for simpler
function (\enquote{Occam's razor}, models with less knots are preferred in the Bayesian modeling, provided they can fit the data equally well),
and (iii) by the CMB constraint on early reionization via $\tau_{15,30}$.

\sh{Based on the FlexKnot samples, we can compute posterior probability distributions for
any derived quantity related to the reionization history. Here we choose to constrain the redshifts of the start ($x_i=0.1$), midpoint ($x_i=0.5$)
and end ($x_i=0.9$) of reionization.
The bounds on these characteristic redshifts are computed in post-processing and this calculation does not affect other parts
of the analysis (such as the derivation of constraints on $\tau$ which we discuss below).}
The redshift constraints are shown in Figure \ref{fig:flexknotxi} (blue error bars) with  the error bars indicating 68\% and 95\% confidence limits
relative to the peak of the posterior distribution, $\hat P$.\footnote{The shift seen in Figure \ref{fig:flexknotxi}, between
the constraints on characteristic EoR points (blue error bars)
and the posterior distribution of reionization histories
(orange contours) \sh{is expected in our analysis.
The orange posterior distribution is subject to the FlexKnot priors (purple),
while the blue error bars are prior-corrected as explained in Section \ref{paragraph:priorcorrection}.
The EoR parameters in Table \ref{tab:resultstable} are each corrected with
respect to a flat prior too.}}
The results are also summarized in Table \ref{tab:resultstable}.
With 100 FRBs we constrain the end,  midpoint and start of reionization to an accuracy of $7\%$, $4\%$ and $11\%$  
respectively (at 68\% confidence level). The corresponding  constraints obtained with 1,000 FRBs are tighter at lower redshifts, i.e. for the end and the  midpoint of reionization, being $5\%$ and $3\%$ respectively. However, the constraint on the start of the EoR is poor ($12\%$) and is   not improved with the sample size. This is because, due to the scarceness of the highest redshift  FRBs in our synthetic data, the uncertainties on the start of reionization are highly asymmetric and the high-redshift tail of the EoR is poorly constrained. 

{It is illustrative to compare the predicted FRB constraints of the EoR to the  present-day state-of-the-art limits. Existing reionization constraints are largely based on quasar and CMB observations. For example,}
 a comprehensive analysis using physical reionization
history models \citep[e.g.][]{Greig2017} shows that  the dark fraction of Lyman $\alpha$ and $\beta$ forest observations,
combined with \textit{Planck} measurements, can constrain the midpoint of reionization to $\sim 12\%$ 
(at 68\% confidence level).
{Comparing to the forecast for FRBs presented in this paper, 
we see that even 100 FRBs at $5<z<15$ could significantly improve   observational constraints on the EoR history,
given the redshift distribution and levels of uncertainties assumed here.
}

\begin{table}
    \centering
    \begin{tabular}{cc@{ }c@{ }cc@{ }c@{ }c}
        \toprule
        Quantity & \multicolumn{3}{c}{100 FRBs} & \multicolumn{3}{c}{{1,000} FRBs} \\
         & $\hat P$ & 68\% & 95\% & $\hat P$ & 68\% & 95\% \\
\midrule
$x_i=0.1$ & $z=7.8$ & $\substack{+0.9\\-0.8}$ & $\substack{+3.8\\-1.1}$ & $z=8.5$ & $\substack{+1.2\\-0.8}$ & $\substack{+4.1\\-1.1}$ \\
\addlinespace[3pt]
$x_i=0.5$ & $z=7.0$ & $\substack{+0.3\\-0.3}$ & $\substack{+0.6\\-0.6}$ & $z=7.2$ & $\substack{+0.1\\-0.3}$ & $\substack{+0.3\\-0.6}$ \\
\addlinespace[3pt]
$x_i=0.9$ & $z=6.1$ & $\substack{+0.5\\-0.4}$ & $\substack{+0.9\\-0.7}$ & $z=6.1$ & $\substack{+0.4\\-0.2}$ & $\substack{+0.6\\-0.5}$ \\
\addlinespace[5pt]
 $\tau$ & $0.0510$ & $\substack{+0.0071\\-0.0036}$ & $\substack{+0.0187\\-0.0058}$ & $0.0548$ & $\substack{+0.0061\\-0.0034}$ & $\substack{+0.0150\\-0.0048}$ \\
\addlinespace[3pt]
$\tau_{\rm w/o\ CMB}$ & $0.0506$ & $\substack{+0.0085\\-0.0042}$ & $\substack{+0.0363\\-0.0053}$ & $0.0537$ & $\substack{+0.0103\\-0.0034}$ & $\substack{+0.0261\\-0.0039}$ \\
\bottomrule
    \end{tabular}
    \caption{FRB constraints on characteristic points of the reionization history: the start, $x_i=0.1$, midpoint, $x_i = 0.5$, and end, $x_i = 0.9$, of the EoR. We also show the {FRB} constraint on the total CMB optical depth {$\tau$} with and without the CMB constraint on $\tau_{15,30}$ \citep{Heinrich2021}. %
    We list the most probable value (i.e.\ location of maximum $\hat P$ of the respective marginalized posterior)
    and the 68\% and 95\% confidence intervals (iso-probability limits enclosing 68\% or 95\% of the
    posterior volume).
    For comparison, we note that the input (fiducial) reionization history used to generate the data has the optical depth of $\tau=0.057$,
    and the start, midpoint and end of reionization at redshifts 10.36, 7.32 and 5.95, respectively, all of which lie within our 95\% confidence intervals.
    }
    \label{tab:resultstable}
\end{table}

\subsection{Optical depth}
\label{subsec:opticaldepth}
\begin{figure}[t]
    \centering
        \includegraphics[width=0.5\textwidth, trim={0 0cm 0 0},clip]{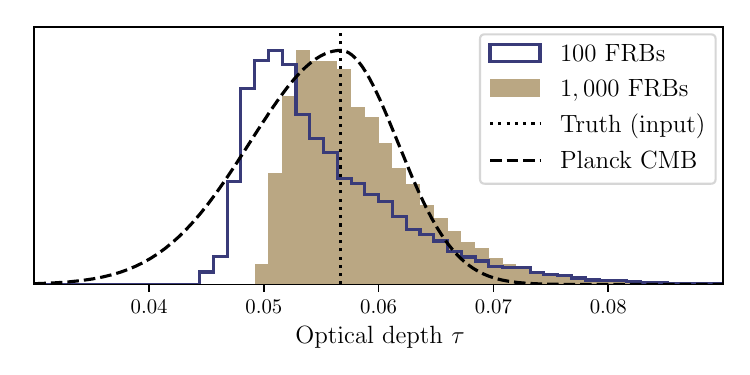}
        \caption{Posterior constraints on the total optical depth $\tau$, marginalized over reionization histories as well as cosmological parameters $\Omega_b \cdot H_0$ and $\Omega_m$.
        Here we show the constraints
        from 100 and 1,000 FRBs, both including the $\tau_{15,30}$ limit.
        We compare the constraints to the full \citetalias{Planck2018} FlexKnot measurement $\tau=0.0504\protect\substack{+0.0050\\-0.0079}$ (68\% confidence, black dashed line). Note that for the ease of comparison we shifted the {\it Planck} posterior distribution to align its mean with our input history. We see that FRBs can significantly improve the lower limit on the total optical depth constraint.
        }
    \label{fig:frb_triangle}
\end{figure}

Using the FRB samples  we can also derive constraints on the optical depth $\tau$,
\sh{ marginalizing over all possible EoR histories and the relevant cosmological parameters.}
We derive the posterior probability distribution for $\tau$ assuming 100 and 1,000 FRBs (shown in Figure \ref{fig:frb_triangle}) and list the detailed limits in Table \ref{tab:resultstable}. We find that 100 and 1,000 FRBs can constrain the optical depth to  $11\%$ and $9\%$ accuracy respectively (at 68\% confidence level; note, however, the asymmetric uncertainties in $\tau$).
The precision is comparable to the optical depth measurement of $\tau=0.0504\substack{+0.0050\\-0.0079}$ (13\% accuracy, at 68\% confidence) derived by the 
\citetalias{Planck2018} collaboration  using a FlexKnot analysis.  However, owing to the asymmetric uncertainties in $\tau$ resulting from our 
FRBs analysis, we find that FRBs provide a much stronger lower limit on $\tau$ than \citetalias{Planck2018}, while their upper bound is weaker due to the low SFR at high redshifts.
This behavior also persists if we remove the $\tau_{15,30}$ constraint (last row of Table \ref{tab:resultstable}), the lower bounds on $\tau$ stay at nearly the same values.

To explore the origin of the asymmetric optical depth constraints, we consider the contributions  of low,
$\tau_{z<10}$,  and high,  $\tau_{z>10}$, redshift to $\tau$ separately.
For the former we find a tight constraint  that improves with the data size,
with the  uncertainty (at 68\% confidence level) decreasing from  5\% for 100 to 2\%  for 1,000 FRBs.\footnote{More precisely the constraints are 
$\tau_{z<10}=0.0508 \substack{+0.0027\\-0.0023}$ for the sample of 100 FRBs and 
$\tau_{z<10}=0.0528 \substack{+0.0011\\-0.0011}$ for  1,000  FRBs.} For the high redshift contribution, $\tau_{z>10}$, we can only obtain upper limits and find  $\tau_{z>10}<0.0049$ for 100 FRBs and $\tau_{z>10}<0.0071$ for  1,000 FRBs at 68\% confidence. These upper limits are compatible with the true value $\tau_{z>10}=0.0024$, 
but do not yield stringent constraints for either data set.
The expected scarcity of the high-redshift FRBs leading to the weak constraints on the high-redshift contribution to $\tau$  is the main limiting factor in precision $\tau$ measurement with FRBs. However,
as this contribution cannot be negative, it barely affects the lower bound on $\tau$ thus allowing to derive stringent lower limits.

{
As mentioned in Section \ref{sec:intro}, the FlexKnot method can
give more accurate FRB EoR constraints than the common fixed-shape
techniques.
As a specific example, we consider the popular tanh parameterization \citep[used e.g. in][]{Zhang2021},
applied to our synthetic data set. Sampling its two parameters (redshift of reionization
$z_{\rm reio}$ and width $\Delta z$) yields a posterior constraint on the optical depth
$\tau_{\rm tanh}=0.0494\substack{+0.0030\\-0.0013}$ (68\% confidence) for 100 FRBs  and
$\tau_{\rm tanh}=0.0514\substack{+0.0014\\-0.0006}$ for 1,000 FRBs.
These inferences systematically underestimate the
true optical depth ($\tau=0.057$) by about $10\%$ at a high statistical
significance: The 100- and 1,000-FRBs tanh posteriors exclude the true value
at $>95\%$ and $>99.7\%$ confidence level, respectively,
while the FlexKnot results for the same data (Table \ref{tab:resultstable}) remain consistent with the input at 68\% confidence level.
The reason for this difference lies in the shape of the tanh history, which does not match that of the input reionization history \citep[from][]{Kulkarni2019}.
In a real-life situation the magnitude of this effect will, of course, depend on how close the actual (unknown) EoR history  is to the assumed parameterization.
There is no way of testing the effect size beforehand and,
thus, we suggest using a free-form method like FlexKnot to
avoid this problem altogether.}

\subsection{Cosmological implications of tight $\tau$ constraints from FRBs}
\begin{figure}[t]
    \centering
        \includegraphics[width=0.5\textwidth]{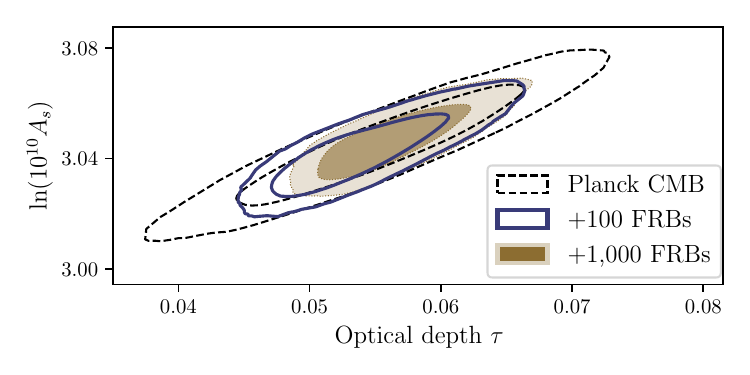}
        \caption{2D posterior constraints on the primordial power spectrum amplitude $A_s$ and the optical depth $\tau$. Compared to the \citetalias{Planck2018} constraints,
        adding a $\tau$ constraints from FRBs significantly decreases the uncertainty in $A_s$. 100 or 1,000 FRBs would allow constraints of $\ln(10^{10} A_s)=3.041\protect\substack{+0.011\\-0.009}$ or $\ln(10^{10}A_s)=3.047\protect\substack{+0.010\\-0.007}$, respectively, compared to the
        \citetalias{Planck2018} constraint of $\ln(10^{10} A_s)=3.043\protect\substack{+0.016\\-0.012}$ ($68\%$ confidence intervals).
        }\label{fig:Astau}
\end{figure}

An accurate optical depth measurement is not only useful as a probe of reionization,
but it will also improve the cosmological {constraints} achievable with other experiments. As an example,
Figure \ref{fig:Astau} shows the \sh{marginalized two-dimensional constraints on $\tau$ and} the initial amplitude of
curvature perturbations, $A_s$, as measured by \citetalias{Planck2018}
(dashed contour lines; from the Planck Legacy Archive\footnote{Full chains for parameter results from \url{https://pla.esac.esa.int/pla/}
(\texttt{COM\_CosmoParams\_fullGrid\_R3.01})
using CMB powerspectra, lensing and Baryonic Acoustic Oscillations.})
which are well known to be degenerate with $\tau$ \citep{Hazra2018}.
An independent optical depth measurement from FRBs allows for tighter constraints on $A_s$  (solid and filled contours in Figure \ref{fig:Astau}). We find that adding the constraints on $\tau$ from 100 or 1,000 FRBs to the \citetalias{Planck2018} data reduces relative error in  $\ln(10^{10} A_s)$ from 0.46\% to 0.32\% and 0.29\% respectively.

Tighter reionization constraints also  improve limits on the sum of neutrino masses, $\sum m_\nu$. Future large scale structure surveys such as
DESI \citep{DESI2016arXivonly}, Euclid \citep{Euclid2018}, and
SKA\footnote{\url{https://www.skatelescope.org/}}, as well as
ground-based CMB telescopes \citep[such as CMB-S4,][]{CMBS42016arXiv},
aim to measure  $\sum m_\nu$ but can only achieve their full potential
in combination with an independent optical depth measurement as demonstrated
by \citet{Brinckmann2019}. They show that CMB-S4 + Euclid can only detect neutrino
masses deviating from zero with a strong lower limit on $\tau$.
Thus an optical depth constraint from FRBs could significantly contribute
to a cosmological detection of the sum of neutrino masses, and in combination
with constraints from large scale CMB polarization measurements
\citep[e.g. from LiteBIRD,][]{LiteBIRD2014}, FRBs can improve the
precision of $\sum m_\nu$ measurements.

\section{Conclusion}
\label{sec:conclusions}

In this work we explore the potential of high redshift FRBs to improve our understanding of the Universe by providing new, independent, constraints on the reionization history. Our analysis is unique in that it is based on  a free-form FlexKnot technique and, thus, the results {are not influenced by assumptions
about the reionization history}.
This is in contrast to  some of the previous works on this topic where  a specific shape of the reionization history was postulated.

Using synthetic FRB data, distributed at $z=5-15$ according to the star formation rate, we show that the full shape of reionization history can be accurately extracted from the dispersion measures of such localized high-redshift events. We find that even a sample of 100 localized FRBs at $z>5$ will allow us to improve reionization constraints beyond the current state-of-the-art CMB bounds and limits derived from high-redshift quasars. A sample of 1,000 FRBs provides tight constraints of $5\%$, $3\%$ and $12\%$  on the  end,  midpoint and start of reionization respectively (corresponding results for 100 localized FRBs are also quoted in Section \ref{sec:results}).

The flexibility of the FlexKnot method allows us to  marginalize over reionization histories and derive model-independent bounds on the CMB optical depth $\tau$. FRBs could improve current
limits on $\tau$ and, through it,
on other cosmological parameters such as the sum of neutrino masses and the amplitude of the power spectrum of primordial fluctuations $A_s$. We show that even 100 FRBs can constrain $\tau$ to an accuracy of $11\%$
(at 68\% confidence level) which is better  than the constraints derived from the Planck FlexKnot analysis ($\sim 13\%$). Owing to the higher numbers of FRBs at lower redshifts, the $\tau$ constraints are asymmetric with a much stronger lower limit of 7\% (at 68\% confidence level). Next, we find that the independent constraints on reionization from 100 FRBs reduces relative error in  $\ln(10^{10} A_s)$ from 0.46\% to  0.32\% when added to the \citetalias{Planck2018} data. The tight lower limit  on $\tau$ will help constraining the sum of neutrino masses from below, potentially contributing
to a cosmological detection of the neutrino mass.

Finally, we note that measuring  the overall amplitude of the $\mathrm{DM}(z)$ relation at high redshifts could, in principle, allow us to constrain other cosmological parameters 
as well, since the amplitude is
proportional to the combination $\Omega_b H_0 \Omega_m^{-0.5}$. However, we do not focus on such constraints here because these parameters
can be constrained by the low-redshift ($z<5$) FRB data to a much greater accuracy \citep{Zhou2014, Hagstotz2021}.

In summary, a successful high-$z$ FRB survey  could substantially advance our understanding of reionization and provide powerful new constrains on cosmological parameters.

\section*{Acknowledgements}

We thank A. Boyle, S. Gratton, M. Haehnelt, W. Handley and L. Hergt for helpful discussions.
{We also thank the anonymous referee for their comments and suggestions that helped improving our paper.}
We acknowledge the usage of the DiRAC HPC.
AF was supported by the Royal Society University Research Fellowship.
NSS was funded by the URF Enhancement Award of AF.
DRL acknowledges support from NSF awards AAG-1616042, OIA-1458952 and PHY-1430284.
SH acknowledges the support of STFC, via the award of a DTP Ph.D. studentship, and the Institute of Astronomy for a maintenance award.

\section*{Tools and data availability:}
In the interest of reproducibility and to facilitate the usage of this method, we make our source 
code and data publicly available \citep[data and codes in][and updated codes on GitHub]{zenodo}\footnote{\url{https://github.com/Stefan-Heimersheim/FlexKnotFRB}}.

We sample the posterior parameter spaces using the \texttt{PolyChord} code \citep{Handley2015a, Handley2015b} and the \texttt{cobaya} wrapper \citep{cobayaASCL, cobayaPaper}.
We visualize these distributions using the analysis codes \texttt{fgivenx} \citep[functional posterior, ][]{Handley2018} and \texttt{anesthetic} \citep[contour plots,][]{Handley2019b}.
The codes make extensive use of the
\texttt{astropy} \citep{astropy2013, astropy2018}, \texttt{scipy} \citep{scipy2020} and \texttt{numpy} \citep{numpy2020} \texttt{python} libraries.


\bibliographystyle{aasjournal}
\bibliography{sample631}{}

\let\jnlstyle=\rm\def\jref#1{{\jnlstyle#1}}\def\aj{\jref{AJ}}
  \def\araa{\jref{ARA\&A}} \def\apj{\jref{ApJ}\ } \def\apjl{\jref{ApJ}\ }
  \def\apjs{\jref{ApJS}} \def\ao{\jref{Appl.~Opt.}} \def\apss{\jref{Ap\&SS}}
  \def\aap{\jref{A\&A}} \def\aapr{\jref{A\&A~Rev.}} \def\aaps{\jref{A\&AS}}
  \def\azh{\jref{AZh}} \def\baas{\jref{BAAS}} \def\jrasc{\jref{JRASC}}
  \def\memras{\jref{MmRAS}} \def\mnras{\jref{MNRAS}\ }
  \def\pra{\jref{Phys.~Rev.~A}\ } \def\prb{\jref{Phys.~Rev.~B}\ }
  \def\prc{\jref{Phys.~Rev.~C}\ } \def\prd{\jref{Phys.~Rev.~D}\ }
  \def\pre{\jref{Phys.~Rev.~E}} \def\prl{\jref{Phys.~Rev.~Lett.}}
  \def\pasp{\jref{PASP}} \def\pasj{\jref{PASJ}} \def\qjras{\jref{QJRAS}}
  \def\skytel{\jref{S\&T}} \def\solphys{\jref{Sol.~Phys.}}
  \def\sovast{\jref{Soviet~Ast.}} \def\ssr{\jref{Space~Sci.~Rev.}}
  \def\zap{\jref{ZAp}} \def\nat{\jref{Nature}\ } \def\iaucirc{\jref{IAU~Circ.}}
  \def\aplett{\jref{Astrophys.~Lett.}}
  \def\apspr{\jref{Astrophys.~Space~Phys.~Res.}}
  \def\bain{\jref{Bull.~Astron.~Inst.~Netherlands}}
  \def\fcp{\jref{Fund.~Cosmic~Phys.}} \def\gca{\jref{Geochim.~Cosmochim.~Acta}}
  \def\grl{\jref{Geophys.~Res.~Lett.}} \def\jcp{\jref{J.~Chem.~Phys.}}
  \def\jgr{\jref{J.~Geophys.~Res.}}
  \def\jqsrt{\jref{J.~Quant.~Spec.~Radiat.~Transf.}}
  \def\memsai{\jref{Mem.~Soc.~Astron.~Italiana}}
  \def\nphysa{\jref{Nucl.~Phys.~A}} \def\physrep{\jref{Phys.~Rep.}}
  \def\physscr{\jref{Phys.~Scr}} \def\planss{\jref{Planet.~Space~Sci.}}
  \def\procspie{\jref{Proc.~SPIE}} \let\astap=\aap \let\apjlett=\apjl
  \let\apjsupp=\apjs \let\applopt=\ao \def\jcap{\jref{JCAP}}
  \let\jnlstyle=\rm\def\jref#1{{\jnlstyle#1}}\def\aj{\jref{AJ}}
  \def\araa{\jref{ARA\&A}} \def\apj{\jref{ApJ}\ } \def\apjl{\jref{ApJ}\ }
  \def\apjs{\jref{ApJS}} \def\ao{\jref{Appl.~Opt.}} \def\apss{\jref{Ap\&SS}}
  \def\aap{\jref{A\&A}} \def\aapr{\jref{A\&A~Rev.}} \def\aaps{\jref{A\&AS}}
  \def\azh{\jref{AZh}} \def\baas{\jref{BAAS}} \def\jrasc{\jref{JRASC}}
  \def\memras{\jref{MmRAS}} \def\mnras{\jref{MNRAS}\ }
  \def\pra{\jref{Phys.~Rev.~A}\ } \def\prb{\jref{Phys.~Rev.~B}\ }
  \def\prc{\jref{Phys.~Rev.~C}\ } \def\prd{\jref{Phys.~Rev.~D}\ }
  \def\pre{\jref{Phys.~Rev.~E}} \def\prl{\jref{Phys.~Rev.~Lett.}}
  \def\pasp{\jref{PASP}} \def\pasj{\jref{PASJ}} \def\qjras{\jref{QJRAS}}
  \def\skytel{\jref{S\&T}} \def\solphys{\jref{Sol.~Phys.}}
  \def\sovast{\jref{Soviet~Ast.}} \def\ssr{\jref{Space~Sci.~Rev.}}
  \def\zap{\jref{ZAp}} \def\nat{\jref{Nature}\ } \def\iaucirc{\jref{IAU~Circ.}}
  \def\aplett{\jref{Astrophys.~Lett.}}
  \def\apspr{\jref{Astrophys.~Space~Phys.~Res.}}
  \def\bain{\jref{Bull.~Astron.~Inst.~Netherlands}}
  \def\fcp{\jref{Fund.~Cosmic~Phys.}} \def\gca{\jref{Geochim.~Cosmochim.~Acta}}
  \def\grl{\jref{Geophys.~Res.~Lett.}} \def\jcp{\jref{J.~Chem.~Phys.}}
  \def\jgr{\jref{J.~Geophys.~Res.}}
  \def\jqsrt{\jref{J.~Quant.~Spec.~Radiat.~Transf.}}
  \def\memsai{\jref{Mem.~Soc.~Astron.~Italiana}}
  \def\nphysa{\jref{Nucl.~Phys.~A}} \def\physrep{\jref{Phys.~Rep.}}
  \def\physscr{\jref{Phys.~Scr}} \def\planss{\jref{Planet.~Space~Sci.}}
  \def\procspie{\jref{Proc.~SPIE}} \let\astap=\aap \let\apjlett=\apjl
  \let\apjsupp=\apjs \let\applopt=\ao \def\jcap{\jref{JCAP}}
\begin{thebibliography}{}
\expandafter\ifx\csname natexlab\endcsname\relax\def\natexlab#1{#1}\fi
\providecommand{\url}[1]{\href{#1}{#1}}
\providecommand{\dodoi}[1]{doi:~\href{http://doi.org/#1}{\nolinkurl{#1}}}
\providecommand{\doeprint}[1]{\href{http://ascl.net/#1}{\nolinkurl{http://ascl.net/#1}}}
\providecommand{\doarXiv}[1]{\href{https://arxiv.org/abs/#1}{\nolinkurl{https://arxiv.org/abs/#1}}}

\bibitem[{{Abazajian} {et~al.}(2016){Abazajian}, {Adshead}, {Ahmed}, {Allen},
  {Alonso}, {Arnold}, {Baccigalupi}, {Bartlett}, {Battaglia}, {Benson},
  {Bischoff}, {Borrill}, {Buza}, {Calabrese}, {Caldwell}, {Carlstrom}, {Chang},
  {Crawford}, {Cyr-Racine}, {De Bernardis}, {de Haan}, {di Serego Alighieri},
  {Dunkley}, {Dvorkin}, {Errard}, {Fabbian}, {Feeney}, {Ferraro}, {Filippini},
  {Flauger}, {Fuller}, {Gluscevic}, {Green}, {Grin}, {Grohs}, {Henning},
  {Hill}, {Hlozek}, {Holder}, {Holzapfel}, {Hu}, {Huffenberger}, {Keskitalo},
  {Knox}, {Kosowsky}, {Kovac}, {Kovetz}, {Kuo}, {Kusaka}, {Le Jeune}, {Lee},
  {Lilley}, {Loverde}, {Madhavacheril}, {Mantz}, {Marsh}, {McMahon},
  {Meerburg}, {Meyers}, {Miller}, {Munoz}, {Nguyen}, {Niemack}, {Peloso},
  {Peloton}, {Pogosian}, {Pryke}, {Raveri}, {Reichardt}, {Rocha}, {Rotti},
  {Schaan}, {Schmittfull}, {Scott}, {Sehgal}, {Shandera}, {Sherwin}, {Smith},
  {Sorbo}, {Starkman}, {Story}, {van Engelen}, {Vieira}, {Watson}, {Whitehorn},
  \& {Kimmy Wu}}]{CMBS42016arXiv}
{Abazajian}, K.~N., {Adshead}, P., {Ahmed}, Z., {et~al.} 2016, arXiv e-prints,
  arXiv:1610.02743.
\newblock \doarXiv{1610.02743}

\bibitem[{{Agarwal} {et~al.}(2020){Agarwal}, {Lorimer}, {Surnis}, {Pei},
  {Karastergiou}, {Golpayegani}, {Werthimer}, {Cobb}, {McLaughlin}, {White},
  {Armour}, {MacMahon}, {Siemion}, \& {Foster}}]{Agarwal:2020}
{Agarwal}, D., {Lorimer}, D.~R., {Surnis}, M.~P., {et~al.} 2020, \mnras, 497,
  352, \dodoi{10.1093/mnras/staa1927}

\bibitem[{{Amendola} {et~al.}(2018){Amendola}, {Appleby}, {Avgoustidis},
  {Bacon}, {Baker}, {Baldi}, {Bartolo}, {Blanchard}, {Bonvin}, {Borgani},
  {Branchini}, {Burrage}, {Camera}, {Carbone}, {Casarini}, {Cropper}, {de
  Rham}, {Dietrich}, {Di Porto}, {Durrer}, {Ealet}, {Ferreira}, {Finelli},
  {Garc{\'\i}a-Bellido}, {Giannantonio}, {Guzzo}, {Heavens}, {Heisenberg},
  {Heymans}, {Hoekstra}, {Hollenstein}, {Holmes}, {Hwang}, {Jahnke},
  {Kitching}, {Koivisto}, {Kunz}, {La Vacca}, {Linder}, {March}, {Marra},
  {Martins}, {Majerotto}, {Markovic}, {Marsh}, {Marulli}, {Massey}, {Mellier},
  {Montanari}, {Mota}, {Nunes}, {Percival}, {Pettorino}, {Porciani},
  {Quercellini}, {Read}, {Rinaldi}, {Sapone}, {Sawicki}, {Scaramella},
  {Skordis}, {Simpson}, {Taylor}, {Thomas}, {Trotta}, {Verde}, {Vernizzi},
  {Vollmer}, {Wang}, {Weller}, \& {Zlosnik}}]{Euclid2018}
{Amendola}, L., {Appleby}, S., {Avgoustidis}, A., {et~al.} 2018, Living Reviews
  in Relativity, 21, 2, \dodoi{10.1007/s41114-017-0010-3}

\bibitem[{{Arcus} {et~al.}(2021){Arcus}, {Macquart}, {Sammons}, {James}, \&
  {Ekers}}]{Arcus2021}
{Arcus}, W.~R., {Macquart}, J.~P., {Sammons}, M.~W., {James}, C.~W., \&
  {Ekers}, R.~D. 2021, \mnras, 501, 5319, \dodoi{10.1093/mnras/staa3948}

\bibitem[{{Astropy Collaboration} {et~al.}(2013){Astropy Collaboration},
  {Robitaille}, {Tollerud}, {Greenfield}, {Droettboom}, {Bray}, {Aldcroft},
  {Davis}, {Ginsburg}, {Price-Whelan}, {Kerzendorf}, {Conley}, {Crighton},
  {Barbary}, {Muna}, {Ferguson}, {Grollier}, {Parikh}, {Nair}, {Unther},
  {Deil}, {Woillez}, {Conseil}, {Kramer}, {Turner}, {Singer}, {Fox}, {Weaver},
  {Zabalza}, {Edwards}, {Azalee Bostroem}, {Burke}, {Casey}, {Crawford},
  {Dencheva}, {Ely}, {Jenness}, {Labrie}, {Lim}, {Pierfederici}, {Pontzen},
  {Ptak}, {Refsdal}, {Servillat}, \& {Streicher}}]{astropy2013}
{Astropy Collaboration}, {Robitaille}, T.~P., {Tollerud}, E.~J., {et~al.} 2013,
  \aap, 558, A33, \dodoi{10.1051/0004-6361/201322068}

\bibitem[{{Astropy Collaboration} {et~al.}(2018){Astropy Collaboration},
  {Price-Whelan}, {Sip{\H{o}}cz}, {G{\"u}nther}, {Lim}, {Crawford}, {Conseil},
  {Shupe}, {Craig}, {Dencheva}, {Ginsburg}, {Vand erPlas}, {Bradley},
  {P{\'e}rez-Su{\'a}rez}, {de Val-Borro}, {Aldcroft}, {Cruz}, {Robitaille},
  {Tollerud}, {Ardelean}, {Babej}, {Bach}, {Bachetti}, {Bakanov}, {Bamford},
  {Barentsen}, {Barmby}, {Baumbach}, {Berry}, {Biscani}, {Boquien}, {Bostroem},
  {Bouma}, {Brammer}, {Bray}, {Breytenbach}, {Buddelmeijer}, {Burke},
  {Calderone}, {Cano Rodr{\'\i}guez}, {Cara}, {Cardoso}, {Cheedella}, {Copin},
  {Corrales}, {Crichton}, {D'Avella}, {Deil}, {Depagne}, {Dietrich}, {Donath},
  {Droettboom}, {Earl}, {Erben}, {Fabbro}, {Ferreira}, {Finethy}, {Fox},
  {Garrison}, {Gibbons}, {Goldstein}, {Gommers}, {Greco}, {Greenfield},
  {Groener}, {Grollier}, {Hagen}, {Hirst}, {Homeier}, {Horton}, {Hosseinzadeh},
  {Hu}, {Hunkeler}, {Ivezi{\'c}}, {Jain}, {Jenness}, {Kanarek}, {Kendrew},
  {Kern}, {Kerzendorf}, {Khvalko}, {King}, {Kirkby}, {Kulkarni}, {Kumar},
  {Lee}, {Lenz}, {Littlefair}, {Ma}, {Macleod}, {Mastropietro}, {McCully},
  {Montagnac}, {Morris}, {Mueller}, {Mumford}, {Muna}, {Murphy}, {Nelson},
  {Nguyen}, {Ninan}, {N{\"o}the}, {Ogaz}, {Oh}, {Parejko}, {Parley}, {Pascual},
  {Patil}, {Patil}, {Plunkett}, {Prochaska}, {Rastogi}, {Reddy Janga},
  {Sabater}, {Sakurikar}, {Seifert}, {Sherbert}, {Sherwood-Taylor}, {Shih},
  {Sick}, {Silbiger}, {Singanamalla}, {Singer}, {Sladen}, {Sooley},
  {Sornarajah}, {Streicher}, {Teuben}, {Thomas}, {Tremblay}, {Turner},
  {Terr{\'o}n}, {van Kerkwijk}, {de la Vega}, {Watkins}, {Weaver}, {Whitmore},
  {Woillez}, {Zabalza}, \& {Astropy Contributors}}]{astropy2018}
{Astropy Collaboration}, {Price-Whelan}, A.~M., {Sip{\H{o}}cz}, B.~M., {et~al.}
  2018, \aj, 156, 123, \dodoi{10.3847/1538-3881/aabc4f}

\bibitem[{{Bannister} {et~al.}(2019){Bannister}, {Deller}, {Phillips},
  {Macquart}, {Prochaska}, {Tejos}, {Ryder}, {Sadler}, {Shannon}, {Simha},
  {Day}, {McQuinn}, {North-Hickey}, {Bhandari}, {Arcus}, {Bennert}, {Burchett},
  {Bouwhuis}, {Dodson}, {Ekers}, {Farah}, {Flynn}, {James}, {Kerr}, {Lenc},
  {Mahony}, {O'Meara}, {Os{\l}owski}, {Qiu}, {Treu}, {U}, {Bateman}, {Bock},
  {Bolton}, {Brown}, {Bunton}, {Chippendale}, {Cooray}, {Cornwell}, {Gupta},
  {Hayman}, {Kesteven}, {Koribalski}, {MacLeod}, {McClure-Griffiths},
  {Neuhold}, {Norris}, {Pilawa}, {Qiao}, {Reynolds}, {Roxby}, {Shimwell},
  {Voronkov}, \& {Wilson}}]{Bannister2019}
{Bannister}, K.~W., {Deller}, A.~T., {Phillips}, C., {et~al.} 2019, Science,
  365, 565, \dodoi{10.1126/science.aaw5903}

\bibitem[{{Batten} {et~al.}(2021){Batten}, {Duffy}, {Wijers}, {Gupta}, {Flynn},
  {Schaye}, \& {Ryan-Weber}}]{Batten2020}
{Batten}, A.~J., {Duffy}, A.~R., {Wijers}, N.~A., {et~al.} 2021, \mnras, 505,
  5356, \dodoi{10.1093/mnras/stab1528}

\bibitem[{{Behroozi} {et~al.}(2019){Behroozi}, {Wechsler}, {Hearin}, \&
  {Conroy}}]{Behroozi2019}
{Behroozi}, P., {Wechsler}, R.~H., {Hearin}, A.~P., \& {Conroy}, C. 2019,
  \mnras, 488, 3143, \dodoi{10.1093/mnras/stz1182}

\bibitem[{{Beniamini} {et~al.}(2021){Beniamini}, {Kumar}, {Ma}, \&
  {Quataert}}]{Beniamini2020}
{Beniamini}, P., {Kumar}, P., {Ma}, X., \& {Quataert}, E. 2021, \mnras, 502,
  5134, \dodoi{10.1093/mnras/stab309}

\bibitem[{{Bhandari} {et~al.}(2018){Bhandari}, {Keane}, {Barr}, {Jameson},
  {Petroff}, {Johnston}, {Bailes}, {Bhat}, {Burgay}, {Burke-Spolaor}, {Caleb},
  {Eatough}, {Flynn}, {Green}, {Jankowski}, {Kramer}, {Krishnan}, {Morello},
  {Possenti}, {Stappers}, {Tiburzi}, {van Straten}, {Andreoni}, {Butterley},
  {Chandra}, {Cooke}, {Corongiu}, {Coward}, {Dhillon}, {Dodson}, {Hardy},
  {Howell}, {Jaroenjittichai}, {Klotz}, {Littlefair}, {Marsh}, {Mickaliger},
  {Muxlow}, {Perrodin}, {Pritchard}, {Sawangwit}, {Terai}, {Tominaga}, {Torne},
  {Totani}, {Trois}, {Turpin}, {Niino}, {Wilson}, {Albert}, {Andr{\'e}},
  {Anghinolfi}, {Anton}, {Ardid}, {Aubert}, {Avgitas}, {Baret},
  {Barrios-Mart{\'\i}}, {Basa}, {Belhorma}, {Bertin}, {Biagi}, {Bormuth},
  {Bourret}, {Bouwhuis}, {Br{\^a}nza{\textcommabelow s}}, {Bruijn}, {Brunner},
  {Busto}, {Capone}, {Caramete}, {Carr}, {Celli}, {Moursli}, {Chiarusi},
  {Circella}, {Coelho}, {Coleiro}, {Coniglione}, {Costantini}, {Coyle},
  {Creusot}, {D{\'\i}az}, {Deschamps}, {De Bonis}, {Distefano}, {Palma},
  {Domi}, {Donzaud}, {Dornic}, {Drouhin}, {Eberl}, {Bojaddaini}, {Khayati},
  {Els{\"a}sser}, {Enzenh{\"o}fer}, {Ettahiri}, {Fassi}, {Felis}, {Fusco},
  {Gay}, {Giordano}, {Glotin}, {Gregoire}, {Gracia-Ruiz}, {Graf}, {Hallmann},
  {van Haren}, {Heijboer}, {Hello}, {Hern{\'a}ndez-Rey}, {H{\"o}{\ss}l},
  {Hofest{\"a}dt}, {Hugon}, {Illuminati}, {James}, {de Jong}, {Jongen},
  {Kadler}, {Kalekin}, {Katz}, {Kie{\ss}ling}, {Kouchner}, {Kreter},
  {Kreykenbohm}, {Kulikovskiy}, {Lachaud}, {Lahmann}, {Lef{\`e}vre}, {Leonora},
  {Loucatos}, {Marcelin}, {Margiotta}, {Marinelli}, {Mart{\'\i}nez-Mora},
  {Mele}, {Melis}, {Michael}, {Migliozzi}, {Moussa}, {Navas}, {Nezri},
  {Organokov}, {P{\v{a}}v{\v{a}}la{\textcommabelow s}}, {Pellegrino},
  {Perrina}, {Piattelli}, {Popa}, {Pradier}, {Quinn}, {Racca}, {Riccobene},
  {S{\'a}nchez-Losa}, {Salda{\~n}a}, {Salvadori}, {Samtleben}, {Sanguineti},
  {Sapienza}, {Sch{\"u}ssler}, {Sieger}, {Spurio}, {Stolarczyk}, {Taiuti},
  {Tayalati}, {Trovato}, {Turpin}, {T{\"o}nnis}, {Vallage}, {Van Elewyck},
  {Versari}, {Vivolo}, {Vizzocca}, {Wilms}, {Zornoza}, \&
  {Z{\'u}{\~n}iga}}]{2018Bhandari}
{Bhandari}, S., {Keane}, E.~F., {Barr}, E.~D., {et~al.} 2018, \mnras, 475,
  1427, \dodoi{10.1093/mnras/stx3074}

\bibitem[{{Bhattacharya} {et~al.}(2021){Bhattacharya}, {Kumar}, \&
  {Linder}}]{Bhattacharya2020}
{Bhattacharya}, M., {Kumar}, P., \& {Linder}, E.~V. 2021, \prd, 103, 103526,
  \dodoi{10.1103/PhysRevD.103.103526}

\bibitem[{{Bochenek} {et~al.}(2020){Bochenek}, {Ravi}, {Belov}, {Hallinan},
  {Kocz}, {Kulkarni}, \& {McKenna}}]{Bochenek_2020}
{Bochenek}, C.~D., {Ravi}, V., {Belov}, K.~V., {et~al.} 2020, \nat, 587, 59,
  \dodoi{10.1038/s41586-020-2872-x}

\bibitem[{{Brinckmann} {et~al.}(2019){Brinckmann}, {Hooper}, {Archidiacono},
  {Lesgourgues}, \& {Sprenger}}]{Brinckmann2019}
{Brinckmann}, T., {Hooper}, D.~C., {Archidiacono}, M., {Lesgourgues}, J., \&
  {Sprenger}, T. 2019, \jcap, 2019, 059, \dodoi{10.1088/1475-7516/2019/01/059}

\bibitem[{{Condon} \& {Ransom}(2016)}]{2016era..book.....C}
{Condon}, J.~J., \& {Ransom}, S.~M. 2016, {Essential Radio Astronomy}

\bibitem[{{Cordes} \& {Lazio}(2002)}]{Cordes_2002arXiv}
{Cordes}, J.~M., \& {Lazio}, T.~J.~W. 2002, arXiv e-prints, astro.
\newblock \doarXiv{astro-ph/0207156}

\bibitem[{{Dai} \& {Xia}(2021)}]{Dai2020}
{Dai}, J.-P., \& {Xia}, J.-Q. 2021, \jcap, 2021, 050,
  \dodoi{10.1088/1475-7516/2021/05/050}

\bibitem[{{Das} {et~al.}(2021){Das}, {Mathur}, {Gupta}, {Nicastro}, \&
  {Krongold}}]{Das2021}
{Das}, S., {Mathur}, S., {Gupta}, A., {Nicastro}, F., \& {Krongold}, Y. 2021,
  \mnras, 500, 655, \dodoi{10.1093/mnras/staa3299}

\bibitem[{{Deng} \& {Zhang}(2014)}]{Deng2014}
{Deng}, W., \& {Zhang}, B. 2014, \apjl, 783, L35,
  \dodoi{10.1088/2041-8205/783/2/L35}

\bibitem[{{DESI Collaboration} {et~al.}(2016){DESI Collaboration}, {Aghamousa},
  {Aguilar}, {Ahlen}, {Alam}, {Allen}, {Allende Prieto}, {Annis}, {Bailey},
  {Balland}, {Ballester}, {Baltay}, {Beaufore}, {Bebek}, {Beers}, {Bell},
  {Bernal}, {Besuner}, {Beutler}, {Blake}, {Bleuler}, {Blomqvist}, {Blum},
  {Bolton}, {Briceno}, {Brooks}, {Brownstein}, {Buckley-Geer}, {Burden},
  {Burtin}, {Busca}, {Cahn}, {Cai}, {Cardiel-Sas}, {Carlberg}, {Carton},
  {Casas}, {Castander}, {Cervantes-Cota}, {Claybaugh}, {Close}, {Coker},
  {Cole}, {Comparat}, {Cooper}, {Cousinou}, {Crocce}, {Cuby}, {Cunningham},
  {Davis}, {Dawson}, {de la Macorra}, {De Vicente}, {Delubac}, {Derwent},
  {Dey}, {Dhungana}, {Ding}, {Doel}, {Duan}, {Ealet}, {Edelstein},
  {Eftekharzadeh}, {Eisenstein}, {Elliott}, {Escoffier}, {Evatt}, {Fagrelius},
  {Fan}, {Fanning}, {Farahi}, {Farihi}, {Favole}, {Feng}, {Fernandez},
  {Findlay}, {Finkbeiner}, {Fitzpatrick}, {Flaugher}, {Flender}, {Font-Ribera},
  {Forero-Romero}, {Fosalba}, {Frenk}, {Fumagalli}, {Gaensicke}, {Gallo},
  {Garcia-Bellido}, {Gaztanaga}, {Pietro Gentile Fusillo}, {Gerard},
  {Gershkovich}, {Giannantonio}, {Gillet}, {Gonzalez-de-Rivera},
  {Gonzalez-Perez}, {Gott}, {Graur}, {Gutierrez}, {Guy}, {Habib}, {Heetderks},
  {Heetderks}, {Heitmann}, {Hellwing}, {Herrera}, {Ho}, {Holland}, {Honscheid},
  {Huff}, {Hutchinson}, {Huterer}, {Hwang}, {Illa Laguna}, {Ishikawa},
  {Jacobs}, {Jeffrey}, {Jelinsky}, {Jennings}, {Jiang}, {Jimenez}, {Johnson},
  {Joyce}, {Jullo}, {Juneau}, {Kama}, {Karcher}, {Karkar}, {Kehoe}, {Kennamer},
  {Kent}, {Kilbinger}, {Kim}, {Kirkby}, {Kisner}, {Kitanidis}, {Kneib},
  {Koposov}, {Kovacs}, {Koyama}, {Kremin}, {Kron}, {Kronig}, {Kueter-Young},
  {Lacey}, {Lafever}, {Lahav}, {Lambert}, {Lampton}, {Landriau}, {Lang},
  {Lauer}, {Le Goff}, {Le Guillou}, {Le Van Suu}, {Lee}, {Lee}, {Leitner},
  {Lesser}, {Levi}, {L'Huillier}, {Li}, {Liang}, {Lin}, {Linder}, {Loebman},
  {Luki{\'c}}, {Ma}, {MacCrann}, {Magneville}, {Makarem}, {Manera}, {Manser},
  {Marshall}, {Martini}, {Massey}, {Matheson}, {McCauley}, {McDonald},
  {McGreer}, {Meisner}, {Metcalfe}, {Miller}, {Miquel}, {Moustakas}, {Myers},
  {Naik}, {Newman}, {Nichol}, {Nicola}, {Nicolati da Costa}, {Nie}, {Niz},
  {Norberg}, {Nord}, {Norman}, {Nugent}, {O'Brien}, {Oh}, {Olsen}, {Padilla},
  {Padmanabhan}, {Padmanabhan}, {Palanque-Delabrouille}, {Palmese},
  {Pappalardo}, {P{\^a}ris}, {Park}, {Patej}, {Peacock}, {Peiris}, {Peng},
  {Percival}, {Perruchot}, {Pieri}, {Pogge}, {Pollack}, {Poppett}, {Prada},
  {Prakash}, {Probst}, {Rabinowitz}, {Raichoor}, {Ree}, {Refregier}, {Regal},
  {Reid}, {Reil}, {Rezaie}, {Rockosi}, {Roe}, {Ronayette}, {Roodman}, {Ross},
  {Ross}, {Rossi}, {Rozo}, {Ruhlmann-Kleider}, {Rykoff}, {Sabiu}, {Samushia},
  {Sanchez}, {Sanchez}, {Schlegel}, {Schneider}, {Schubnell}, {Secroun},
  {Seljak}, {Seo}, {Serrano}, {Shafieloo}, {Shan}, {Sharples}, {Sholl},
  {Shourt}, {Silber}, {Silva}, {Sirk}, {Slosar}, {Smith}, {Smoot}, {Som},
  {Song}, {Sprayberry}, {Staten}, {Stefanik}, {Tarle}, {Sien Tie}, {Tinker},
  {Tojeiro}, {Valdes}, {Valenzuela}, {Valluri}, {Vargas-Magana}, {Verde},
  {Walker}, {Wang}, {Wang}, {Weaver}, {Weaverdyck}, {Wechsler}, {Weinberg},
  {White}, {Yang}, {Yeche}, {Zhang}, {Zhao}, {Zheng}, {Zhou}, {Zhou}, {Zhu},
  {Zou}, \& {Zu}}]{DESI2016arXivonly}
{DESI Collaboration}, {Aghamousa}, A., {Aguilar}, J., {et~al.} 2016, arXiv
  e-prints, arXiv:1611.00036.
\newblock \doarXiv{1611.00036}

\bibitem[{{Fialkov} \& {Loeb}(2016)}]{Fialkov_2016}
{Fialkov}, A., \& {Loeb}, A. 2016, \jcap, 2016, 004,
  \dodoi{10.1088/1475-7516/2016/05/004}

\bibitem[{{Fialkov} \& {Loeb}(2017)}]{Fialkov:2017}
---. 2017, \apjl, 846, L27, \dodoi{10.3847/2041-8213/aa8905}

\bibitem[{{Gao} {et~al.}(2014){Gao}, {Li}, \& {Zhang}}]{Gao2014}
{Gao}, H., {Li}, Z., \& {Zhang}, B. 2014, \apj, 788, 189,
  \dodoi{10.1088/0004-637X/788/2/189}

\bibitem[{{Greig} \& {Mesinger}(2017)}]{Greig2017}
{Greig}, B., \& {Mesinger}, A. 2017, \mnras, 465, 4838,
  \dodoi{10.1093/mnras/stw3026}

\bibitem[{{Hagstotz} {et~al.}(2022){Hagstotz}, {Reischke}, \&
  {Lilow}}]{Hagstotz2021}
{Hagstotz}, S., {Reischke}, R., \& {Lilow}, R. 2022, \mnras, 511, 662,
  \dodoi{10.1093/mnras/stac077}

\bibitem[{{Handley}(2018)}]{Handley2018}
{Handley}, W. 2018, The Journal of Open Source Software, 3, 849,
  \dodoi{10.21105/joss.00849}

\bibitem[{{Handley}(2019)}]{Handley2019b}
---. 2019, The Journal of Open Source Software, 4, 1414,
  \dodoi{10.21105/joss.01414}

\bibitem[{{Handley} \& {Millea}(2019)}]{Handley2019}
{Handley}, W., \& {Millea}, M. 2019, Entropy, 21, 272,
  \dodoi{10.3390/e21030272}

\bibitem[{{Handley} {et~al.}(2015{\natexlab{a}}){Handley}, {Hobson}, \&
  {Lasenby}}]{Handley2015a}
{Handley}, W.~J., {Hobson}, M.~P., \& {Lasenby}, A.~N. 2015{\natexlab{a}},
  \mnras, 450, L61, \dodoi{10.1093/mnrasl/slv047}

\bibitem[{{Handley} {et~al.}(2015{\natexlab{b}}){Handley}, {Hobson}, \&
  {Lasenby}}]{Handley2015b}
---. 2015{\natexlab{b}}, \mnras, 453, 4384, \dodoi{10.1093/mnras/stv1911}

\bibitem[{{Handley} {et~al.}(2019){Handley}, {Lasenby}, {Peiris}, \&
  {Hobson}}]{Handley2019Infl}
{Handley}, W.~J., {Lasenby}, A.~N., {Peiris}, H.~V., \& {Hobson}, M.~P. 2019,
  \prd, 100, 103511, \dodoi{10.1103/PhysRevD.100.103511}

\bibitem[{Harris {et~al.}(2020)Harris, Millman, van~der Walt, Gommers,
  Virtanen, Cournapeau, Wieser, Taylor, Berg, Smith, Kern, Picus, Hoyer, van
  Kerkwijk, Brett, Haldane, del R{\'{i}}o, Wiebe, Peterson,
  G{\'{e}}rard-Marchant, Sheppard, Reddy, Weckesser, Abbasi, Gohlke, \&
  Oliphant}]{numpy2020}
Harris, C.~R., Millman, K.~J., van~der Walt, S.~J., {et~al.} 2020, Nature, 585,
  357, \dodoi{10.1038/s41586-020-2649-2}

\bibitem[{{Hashimoto} {et~al.}(2018){Hashimoto}, {Laporte}, {Mawatari},
  {Ellis}, {Inoue}, {Zackrisson}, {Roberts-Borsani}, {Zheng}, {Tamura},
  {Bauer}, {Fletcher}, {Harikane}, {Hatsukade}, {Hayatsu}, {Matsuda}, {Matsuo},
  {Okamoto}, {Ouchi}, {Pell{\'o}}, {Rydberg}, {Shimizu}, {Taniguchi},
  {Umehata}, \& {Yoshida}}]{Hashimoto:2018}
{Hashimoto}, T., {Laporte}, N., {Mawatari}, K., {et~al.} 2018, \nat, 557, 392,
  \dodoi{10.1038/s41586-018-0117-z}

\bibitem[{{Hashimoto} {et~al.}(2020{\natexlab{a}}){Hashimoto}, {Goto}, {On},
  {Lu}, {Santos}, {Ho}, {Wang}, {Kim}, \& {Hsiao}}]{Hashimoto2008.00007}
{Hashimoto}, T., {Goto}, T., {On}, A. Y.~L., {et~al.} 2020{\natexlab{a}},
  \mnras, 497, 4107, \dodoi{10.1093/mnras/staa2238}

\bibitem[{{Hashimoto} {et~al.}(2020{\natexlab{b}}){Hashimoto}, {Goto}, {On},
  {Lu}, {Santos}, {Ho}, {Kim}, {Wang}, \& {Hsiao}}]{Hashimoto2020SFR}
---. 2020{\natexlab{b}}, \mnras, 498, 3927, \dodoi{10.1093/mnras/staa2490}

\bibitem[{{Hashimoto} {et~al.}(2021){Hashimoto}, {Goto}, {Lu}, {On}, {Santos},
  {Kim}, {Kilerci Eser}, {Ho}, {Hsiao}, \& {Lin}}]{HashimotoReioH2021}
{Hashimoto}, T., {Goto}, T., {Lu}, T.-Y., {et~al.} 2021, \mnras, 502, 2346,
  \dodoi{10.1093/mnras/stab186}

\bibitem[{{Hazra} {et~al.}(2018){Hazra}, {Paoletti}, {Finelli}, \&
  {Smoot}}]{Hazra2018}
{Hazra}, D.~K., {Paoletti}, D., {Finelli}, F., \& {Smoot}, G.~F. 2018, \jcap,
  2018, 016, \dodoi{10.1088/1475-7516/2018/09/016}

\bibitem[{{Hazra} \& {Smoot}(2017)}]{Hazra2017}
{Hazra}, D.~K., \& {Smoot}, G.~F. 2017, \jcap, 2017, 028,
  \dodoi{10.1088/1475-7516/2017/11/028}

\bibitem[{Heimersheim {et~al.}(2022)Heimersheim, Sartorio, Fialkov, \&
  Lorimer}]{zenodo}
Heimersheim, S., Sartorio, N.~S., Fialkov, A., \& Lorimer, D.~R. 2022, {Data
  and Code for "What it Takes to Measure Reionization with Fast Radio Bursts"},
   Zenodo, \dodoi{10.5281/zenodo.6542596}

\bibitem[{{Heinrich} \& {Hu}(2018)}]{Heinrich2018}
{Heinrich}, C., \& {Hu}, W. 2018, \prd, 98, 063514,
  \dodoi{10.1103/PhysRevD.98.063514}

\bibitem[{{Heinrich} \& {Hu}(2021)}]{Heinrich2021}
---. 2021, \prd, 104, 063505, \dodoi{10.1103/PhysRevD.104.063505}

\bibitem[{{Hu} \& {Holder}(2003)}]{Hu2003}
{Hu}, W., \& {Holder}, G.~P. 2003, \prd, 68, 023001,
  \dodoi{10.1103/PhysRevD.68.023001}

\bibitem[{{James} {et~al.}(2022){James}, {Prochaska}, {Macquart},
  {North-Hickey}, {Bannister}, \& {Dunning}}]{James:2021}
{James}, C.~W., {Prochaska}, J.~X., {Macquart}, J.~P., {et~al.} 2022, \mnras,
  510, L18, \dodoi{10.1093/mnrasl/slab117}

\bibitem[{{Jaroszynski}(2019)}]{Jaroszynski2019}
{Jaroszynski}, M. 2019, \mnras, 484, 1637, \dodoi{10.1093/mnras/sty3529}

\bibitem[{{Jaroszy{\'n}ski}(2020)}]{Jaroszynski2020}
{Jaroszy{\'n}ski}, M. 2020, \actaa, 70, 87, \dodoi{10.32023/0001-5237/70.2.1}

\bibitem[{{Keating} \& {Pen}(2020)}]{Keating2020}
{Keating}, L.~C., \& {Pen}, U.-L. 2020, \mnras, 496, L106,
  \dodoi{10.1093/mnrasl/slaa095}

\bibitem[{{Klessen}(2019)}]{Klessen:2019}
{Klessen}, R. 2019, {Formation of the first stars}, ed. M.~{Latif} \&
  D.~{Schleicher}, 67--97, \dodoi{10.1142/9789813227958_0004}

\bibitem[{{Kulkarni} {et~al.}(2019){Kulkarni}, {Keating}, {Haehnelt}, {Bosman},
  {Puchwein}, {Chardin}, \& {Aubert}}]{Kulkarni2019}
{Kulkarni}, G., {Keating}, L.~C., {Haehnelt}, M.~G., {et~al.} 2019, \mnras,
  485, L24, \dodoi{10.1093/mnrasl/slz025}

\bibitem[{{Kulkarni}(2020)}]{Kulkarni2020arXiv}
{Kulkarni}, S.~R. 2020, arXiv e-prints, arXiv:2007.02886.
\newblock \doarXiv{2007.02886}

\bibitem[{{Kumar} \& {Linder}(2019)}]{Kumar_2019}
{Kumar}, P., \& {Linder}, E.~V. 2019, \prd, 100, 083533,
  \dodoi{10.1103/PhysRevD.100.083533}

\bibitem[{{Linder}(2020)}]{Linder2020}
{Linder}, E.~V. 2020, \prd, 101, 103019, \dodoi{10.1103/PhysRevD.101.103019}

\bibitem[{{Lorimer} {et~al.}(2007){Lorimer}, {Bailes}, {McLaughlin},
  {Narkevic}, \& {Crawford}}]{Lorimer_2007}
{Lorimer}, D.~R., {Bailes}, M., {McLaughlin}, M.~A., {Narkevic}, D.~J., \&
  {Crawford}, F. 2007, Science, 318, 777, \dodoi{10.1126/science.1147532}

\bibitem[{{Lorimer} {et~al.}(2013){Lorimer}, {Karastergiou}, {McLaughlin}, \&
  {Johnston}}]{Lorimer2013MNRAS.436L...5L}
{Lorimer}, D.~R., {Karastergiou}, A., {McLaughlin}, M.~A., \& {Johnston}, S.
  2013, \mnras, 436, L5, \dodoi{10.1093/mnrasl/slt098}

\bibitem[{{Macquart} {et~al.}(2020){Macquart}, {Prochaska}, {McQuinn},
  {Bannister}, {Bhandari}, {Day}, {Deller}, {Ekers}, {James}, {Marnoch},
  {Os{\l}owski}, {Phillips}, {Ryder}, {Scott}, {Shannon}, \&
  {Tejos}}]{Macquart_2020}
{Macquart}, J.~P., {Prochaska}, J.~X., {McQuinn}, M., {et~al.} 2020, \nat, 581,
  391, \dodoi{10.1038/s41586-020-2300-2}

\bibitem[{Madhavacheril {et~al.}(2019)Madhavacheril, Battaglia, Smith, \&
  Sievers}]{Madhavacheril2019}
Madhavacheril, M.~S., Battaglia, N., Smith, K.~M., \& Sievers, J.~L. 2019,
  Phys. Rev. D, 100, 103532, \dodoi{10.1103/PhysRevD.100.103532}

\bibitem[{{Matsumura} {et~al.}(2014){Matsumura}, {Akiba}, {Borrill}, {Chinone},
  {Dobbs}, {Fuke}, {Ghribi}, {Hasegawa}, {Hattori}, {Hattori}, {Hazumi},
  {Holzapfel}, {Inoue}, {Ishidoshiro}, {Ishino}, {Ishitsuka}, {Karatsu},
  {Katayama}, {Kawano}, {Kibayashi}, {Kibe}, {Kimura}, {Kimura}, {Koga},
  {Kozu}, {Komatsu}, {Lee}, {Matsuhara}, {Mima}, {Mitsuda}, {Mizukami},
  {Morii}, {Morishima}, {Murayama}, {Nagai}, {Nagata}, {Nakamura}, {Naruse},
  {Natsume}, {Nishibori}, {Nishino}, {Noda}, {Noguchi}, {Ogawa}, {Oguri},
  {Ohta}, {Otani}, {Richards}, {Sakai}, {Sato}, {Sato}, {Sekimoto}, {Shimizu},
  {Shinozaki}, {Sugita}, {Suzuki}, {Suzuki}, {Tajima}, {Takada}, {Takakura},
  {Takei}, {Tomaru}, {Uzawa}, {Wada}, {Watanabe}, {Yoshida}, {Yamasaki},
  {Yoshida}, \& {Yotsumoto}}]{LiteBIRD2014}
{Matsumura}, T., {Akiba}, Y., {Borrill}, J., {et~al.} 2014, Journal of Low
  Temperature Physics, 176, 733, \dodoi{10.1007/s10909-013-0996-1}

\bibitem[{{McGreer} {et~al.}(2015){McGreer}, {Mesinger}, \&
  {D'Odorico}}]{McGreer2015}
{McGreer}, I.~D., {Mesinger}, A., \& {D'Odorico}, V. 2015, \mnras, 447, 499,
  \dodoi{10.1093/mnras/stu2449}

\bibitem[{{Millea} \& {Bouchet}(2018)}]{Millea2018}
{Millea}, M., \& {Bouchet}, F. 2018, \aap, 617, A96,
  \dodoi{10.1051/0004-6361/201833288}

\bibitem[{{Nan} {et~al.}(2011){Nan}, {Li}, {Jin}, {Wang}, {Zhu}, {Zhu},
  {Zhang}, {Yue}, \& {Qian}}]{Nan2011}
{Nan}, R., {Li}, D., {Jin}, C., {et~al.} 2011, International Journal of Modern
  Physics D, 20, 989, \dodoi{10.1142/S0218271811019335}

\bibitem[{{Naoz} {et~al.}(2006){Naoz}, {Noter}, \& {Barkana}}]{Naoz:2006}
{Naoz}, S., {Noter}, S., \& {Barkana}, R. 2006, \mnras, 373, L98,
  \dodoi{10.1111/j.1745-3933.2006.00251.x}

\bibitem[{{Oesch} {et~al.}(2016){Oesch}, {Brammer}, {van Dokkum},
  {Illingworth}, {Bouwens}, {Labb{\'e}}, {Franx}, {Momcheva}, {Ashby}, {Fazio},
  {Gonzalez}, {Holden}, {Magee}, {Skelton}, {Smit}, {Spitler}, {Trenti}, \&
  {Willner}}]{Oesch:2016}
{Oesch}, P.~A., {Brammer}, G., {van Dokkum}, P.~G., {et~al.} 2016, \apj, 819,
  129, \dodoi{10.3847/0004-637X/819/2/129}

\bibitem[{{Pagano} \& {Fronenberg}(2021)}]{Pagano2021}
{Pagano}, M., \& {Fronenberg}, H. 2021, \mnras, 505, 2195,
  \dodoi{10.1093/mnras/stab1438}

\bibitem[{{Petroff} {et~al.}(2016){Petroff}, {Barr}, {Jameson}, {Keane},
  {Bailes}, {Kramer}, {Morello}, {Tabbara}, \& {van Straten}}]{Petroff_2016}
{Petroff}, E., {Barr}, E.~D., {Jameson}, A., {et~al.} 2016, \pasa, 33, e045,
  \dodoi{10.1017/pasa.2016.35}

\bibitem[{{Planck Collaboration} {et~al.}(2020){Planck Collaboration},
  {Aghanim}, {Akrami}, {Ashdown}, {Aumont}, {Baccigalupi}, {Ballardini},
  {Banday}, {Barreiro}, {Bartolo}, {Basak}, {Battye}, {Benabed}, {Bernard},
  {Bersanelli}, {Bielewicz}, {Bock}, {Bond}, {Borrill}, {Bouchet}, {Boulanger},
  {Bucher}, {Burigana}, {Butler}, {Calabrese}, {Cardoso}, {Carron},
  {Challinor}, {Chiang}, {Chluba}, {Colombo}, {Combet}, {Contreras}, {Crill},
  {Cuttaia}, {de Bernardis}, {de Zotti}, {Delabrouille}, {Delouis}, {Di
  Valentino}, {Diego}, {Dor{\'e}}, {Douspis}, {Ducout}, {Dupac}, {Dusini},
  {Efstathiou}, {Elsner}, {En{\ss}lin}, {Eriksen}, {Fantaye}, {Farhang},
  {Fergusson}, {Fernandez-Cobos}, {Finelli}, {Forastieri}, {Frailis},
  {Fraisse}, {Franceschi}, {Frolov}, {Galeotta}, {Galli}, {Ganga},
  {G{\'e}nova-Santos}, {Gerbino}, {Ghosh}, {Gonz{\'a}lez-Nuevo}, {G{\'o}rski},
  {Gratton}, {Gruppuso}, {Gudmundsson}, {Hamann}, {Handley}, {Hansen},
  {Herranz}, {Hildebrandt}, {Hivon}, {Huang}, {Jaffe}, {Jones}, {Karakci},
  {Keih{\"a}nen}, {Keskitalo}, {Kiiveri}, {Kim}, {Kisner}, {Knox},
  {Krachmalnicoff}, {Kunz}, {Kurki-Suonio}, {Lagache}, {Lamarre}, {Lasenby},
  {Lattanzi}, {Lawrence}, {Le Jeune}, {Lemos}, {Lesgourgues}, {Levrier},
  {Lewis}, {Liguori}, {Lilje}, {Lilley}, {Lindholm}, {L{\'o}pez-Caniego},
  {Lubin}, {Ma}, {Mac{\'\i}as-P{\'e}rez}, {Maggio}, {Maino}, {Mandolesi},
  {Mangilli}, {Marcos-Caballero}, {Maris}, {Martin}, {Martinelli},
  {Mart{\'\i}nez-Gonz{\'a}lez}, {Matarrese}, {Mauri}, {McEwen}, {Meinhold},
  {Melchiorri}, {Mennella}, {Migliaccio}, {Millea}, {Mitra},
  {Miville-Desch{\^e}nes}, {Molinari}, {Montier}, {Morgante}, {Moss}, {Natoli},
  {N{\o}rgaard-Nielsen}, {Pagano}, {Paoletti}, {Partridge}, {Patanchon},
  {Peiris}, {Perrotta}, {Pettorino}, {Piacentini}, {Polastri}, {Polenta},
  {Puget}, {Rachen}, {Reinecke}, {Remazeilles}, {Renzi}, {Rocha}, {Rosset},
  {Roudier}, {Rubi{\~n}o-Mart{\'\i}n}, {Ruiz-Granados}, {Salvati}, {Sandri},
  {Savelainen}, {Scott}, {Shellard}, {Sirignano}, {Sirri}, {Spencer},
  {Sunyaev}, {Suur-Uski}, {Tauber}, {Tavagnacco}, {Tenti}, {Toffolatti},
  {Tomasi}, {Trombetti}, {Valenziano}, {Valiviita}, {Van Tent}, {Vibert},
  {Vielva}, {Villa}, {Vittorio}, {Wandelt}, {Wehus}, {White}, {White},
  {Zacchei}, \& {Zonca}}]{Planck2018}
{Planck Collaboration}, {Aghanim}, N., {Akrami}, Y., {et~al.} 2020, \aap, 641,
  A6, \dodoi{10.1051/0004-6361/201833910}

\bibitem[{{Popov} \& {Postnov}(2010)}]{Popov_2010arXiv}
{Popov}, S.~B., \& {Postnov}, K.~A. 2010, in Evolution of Cosmic Objects
  through their Physical Activity, ed. H.~A. {Harutyunian}, A.~M. {Mickaelian},
  \& Y.~{Terzian}, 129--132.
\newblock \doarXiv{0710.2006}

\bibitem[{{Rafiei-Ravandi} {et~al.}(2021){Rafiei-Ravandi}, {Smith}, {Li},
  {Masui}, {Josephy}, {Dobbs}, {Lang}, {Bhardwaj}, {Patel}, {Bandura},
  {Berger}, {Boyle}, {Brar}, {Breitman}, {Cassanelli}, {Chawla}, {Adam Dong},
  {Fonseca}, {Gaensler}, {Giri}, {Good}, {Halpern}, {Kaczmarek}, {Kaspi},
  {Leung}, {Lin}, {Mena-Parra}, {Meyers}, {Michilli}, {M{\"u}nchmeyer}, {Ng},
  {Petroff}, {Pleunis}, {Rahman}, {Sanghavi}, {Scholz}, {Shin}, {Stairs},
  {Tendulkar}, {Vanderlinde}, \& {Zwaniga}}]{CHIMEFRBCatalog1}
{Rafiei-Ravandi}, M., {Smith}, K.~M., {Li}, D., {et~al.} 2021, \apj, 922, 42,
  \dodoi{10.3847/1538-4357/ac1dab}

\bibitem[{{Skilling}(2004)}]{Skilling2004}
{Skilling}, J. 2004, in American Institute of Physics Conference Series, Vol.
  735, Bayesian Inference and Maximum Entropy Methods in Science and
  Engineering: 24th International Workshop on Bayesian Inference and Maximum
  Entropy Methods in Science and Engineering, ed. R.~{Fischer}, R.~{Preuss}, \&
  U.~V. {Toussaint}, 395--405, \dodoi{10.1063/1.1835238}

\bibitem[{{Takahashi} {et~al.}(2021){Takahashi}, {Ioka}, {Mori}, \&
  {Funahashi}}]{Takahashi2020}
{Takahashi}, R., {Ioka}, K., {Mori}, A., \& {Funahashi}, K. 2021, \mnras, 502,
  2615, \dodoi{10.1093/mnras/stab170}

\bibitem[{{Tendulkar} {et~al.}(2017){Tendulkar}, {Bassa}, {Cordes}, {Bower},
  {Law}, {Chatterjee}, {Adams}, {Bogdanov}, {Burke-Spolaor}, {Butler},
  {Demorest}, {Hessels}, {Kaspi}, {Lazio}, {Maddox}, {Marcote}, {McLaughlin},
  {Paragi}, {Ransom}, {Scholz}, {Seymour}, {Spitler}, {van Langevelde}, \&
  {Wharton}}]{Tendulkar2017}
{Tendulkar}, S.~P., {Bassa}, C.~G., {Cordes}, J.~M., {et~al.} 2017, \apjl, 834,
  L7, \dodoi{10.3847/2041-8213/834/2/L7}

\bibitem[{{The CHIME/FRB Collaboration} {et~al.}(2020){The CHIME/FRB
  Collaboration}, Andersen, {Band ura}, {Bhardwaj}, {Bij}, {Boyce}, {Boyle},
  {Brar}, {Cassanelli}, {Chawla}, {Chen}, {Cliche}, {Cook}, {Cubranic},
  {Curtin}, {Denman}, {Dobbs}, {Dong}, {Fand ino}, {Fonseca}, {Gaensler},
  {Giri}, {Good}, {Halpern}, {Hill}, {Hinshaw}, {H{\"o}fer}, {Josephy},
  {Kania}, {Kaspi}, {Land ecker}, {Leung}, {Li}, {Lin}, {Masui}, {McKinven},
  {Mena-Parra}, {Merryfield}, {Meyers}, {Michilli}, {Milutinovic},
  {Mirhosseini}, {M{\"u}nchmeyer}, {Naidu}, {Newburgh}, {Ng}, {Patel}, {Pen},
  {Pinsonneault-Marotte}, {Pleunis}, {Quine}, {Rafiei-Ravandi}, {Rahman},
  {Ransom}, {Renard}, {Sanghavi}, {Scholz}, {Shaw}, {Shin}, {Siegel}, {Singh},
  {Smegal}, {Smith}, {Stairs}, {Tan}, {Tendulkar}, {Tretyakov}, {Vanderlinde},
  {Wang}, {Wulf}, \& {Zwaniga}}]{Chime_2020}
{The CHIME/FRB Collaboration}, Andersen, B. {\^A}.~C., {Band ura}, K.
  {\^A}.~M., {et~al.} 2020, \nat, 587, 54, \dodoi{10.1038/s41586-020-2863-y}

\bibitem[{{Torrado} \& {Lewis}(2019)}]{cobayaASCL}
{Torrado}, J., \& {Lewis}, A. 2019, {Cobaya: Bayesian analysis in cosmology}.
\newblock \doeprint{1910.019}

\bibitem[{{Torrado} \& {Lewis}(2021)}]{cobayaPaper}
---. 2021, \jcap, 2021, 057, \dodoi{10.1088/1475-7516/2021/05/057}

\bibitem[{{V{\'a}zquez} {et~al.}(2012){V{\'a}zquez}, {Bridges}, {Hobson}, \&
  {Lasenby}}]{Vazquez2012}
{V{\'a}zquez}, J.~A., {Bridges}, M., {Hobson}, M.~P., \& {Lasenby}, A.~N. 2012,
  \jcap, 2012, 006, \dodoi{10.1088/1475-7516/2012/06/006}

\bibitem[{Virtanen {et~al.}(2020)Virtanen, Gommers, Oliphant, Haberland, Reddy,
  Cournapeau, Burovski, Peterson, Weckesser, Bright, {van der Walt}, Brett,
  Wilson, Millman, Mayorov, Nelson, Jones, Kern, Larson, Carey, Polat, Feng,
  Moore, {VanderPlas}, Laxalde, Perktold, Cimrman, Henriksen, Quintero, Harris,
  Archibald, Ribeiro, Pedregosa, {van Mulbregt}, \& {SciPy 1.0
  Contributors}}]{scipy2020}
Virtanen, P., Gommers, R., Oliphant, T.~E., {et~al.} 2020, Nature Methods, 17,
  261, \dodoi{10.1038/s41592-019-0686-2}

\bibitem[{{Wadiasingh} {et~al.}(2020){Wadiasingh}, {Beniamini}, {Timokhin},
  {Baring}, {van der Horst}, {Harding}, \& {Kazanas}}]{Wadiasingh_2020}
{Wadiasingh}, Z., {Beniamini}, P., {Timokhin}, A., {et~al.} 2020, \apj, 891,
  82, \dodoi{10.3847/1538-4357/ab6d69}

\bibitem[{{Walters} {et~al.}(2019){Walters}, {Ma}, {Sievers}, \&
  {Weltman}}]{Walters2019}
{Walters}, A., {Ma}, Y.-Z., {Sievers}, J., \& {Weltman}, A. 2019, \prd, 100,
  103519, \dodoi{10.1103/PhysRevD.100.103519}

\bibitem[{{Walters} {et~al.}(2018){Walters}, {Weltman}, {Gaensler}, {Ma}, \&
  {Witzemann}}]{Walters2018}
{Walters}, A., {Weltman}, A., {Gaensler}, B.~M., {Ma}, Y.-Z., \& {Witzemann},
  A. 2018, \apj, 856, 65, \dodoi{10.3847/1538-4357/aaaf6b}

\bibitem[{{Wu} {et~al.}(2020){Wu}, {Yu}, \& {Wang}}]{Wu2020}
{Wu}, Q., {Yu}, H., \& {Wang}, F.~Y. 2020, \apj, 895, 33,
  \dodoi{10.3847/1538-4357/ab88d2}

\bibitem[{{Xu} \& {Han}(2015)}]{Xu_2015}
{Xu}, J., \& {Han}, J.~L. 2015, Research in Astronomy and Astrophysics, 15,
  1629, \dodoi{10.1088/1674-4527/15/10/002}

\bibitem[{{Yamasaki} \& {Totani}(2020)}]{Yamasaki2020}
{Yamasaki}, S., \& {Totani}, T. 2020, \apj, 888, 105,
  \dodoi{10.3847/1538-4357/ab58c4}

\bibitem[{{Yang} \& {Zhang}(2016)}]{Yang_2016}
{Yang}, Y.-P., \& {Zhang}, B. 2016, \apjl, 830, L31,
  \dodoi{10.3847/2041-8205/830/2/L31}

\bibitem[{{Yao} {et~al.}(2017){Yao}, {Manchester}, \& {Wang}}]{Yao2017}
{Yao}, J.~M., {Manchester}, R.~N., \& {Wang}, N. 2017, \apj, 835, 29,
  \dodoi{10.3847/1538-4357/835/1/29}

\bibitem[{{Zhang}(2018)}]{Zhang:2018}
{Zhang}, B. 2018, \apjl, 867, L21, \dodoi{10.3847/2041-8213/aae8e3}

\bibitem[{{Zhang} {et~al.}(2021){Zhang}, {Yan}, {Li}, {Zhang}, \&
  {Wang}}]{Zhang2021}
{Zhang}, Z.~J., {Yan}, K., {Li}, C.~M., {Zhang}, G.~Q., \& {Wang}, F.~Y. 2021,
  \apj, 906, 49, \dodoi{10.3847/1538-4357/abceb9}

\bibitem[{{Zheng} {et~al.}(2014){Zheng}, {Ofek}, {Kulkarni}, {Neill}, \&
  {Juric}}]{Zheng2014}
{Zheng}, Z., {Ofek}, E.~O., {Kulkarni}, S.~R., {Neill}, J.~D., \& {Juric}, M.
  2014, \apj, 797, 71, \dodoi{10.1088/0004-637X/797/1/71}

\bibitem[{{Zhou} {et~al.}(2014){Zhou}, {Li}, {Wang}, {Fan}, \&
  {Wei}}]{Zhou2014}
{Zhou}, B., {Li}, X., {Wang}, T., {Fan}, Y.-Z., \& {Wei}, D.-M. 2014, \prd, 89,
  107303, \dodoi{10.1103/PhysRevD.89.107303}

\end{thebibliography}



\end{document}